\documentclass[sigconf]{acmart}

\usepackage[utf8]{inputenc}
\usepackage{xcolor}
\usepackage{color, colortbl}
\usepackage{graphicx,latexsym,subfigure}
\usepackage{multicol}
\usepackage{multirow}
\usepackage{algorithm}
\usepackage{algorithmic}
\usepackage{booktabs} 
\usepackage{amsmath,amssymb,amsfonts}
\usepackage{threeparttable}
\usepackage{enumitem}

\newcommand{\tool}{\textsc{FairPick}}

\newcommand{\Target}{\mathcal{H}}

\newcommand{\MIA}{\mathcal{M}}

\newcommand{\etal}{\mbox{\emph{et al.\ }}}
\newcommand{\eg}{\mbox{\emph{e.g.\ }}}

\newenvironment{ditemize}{
\begin{list}{{\bf $\bullet$}}{
\setlength{\itemsep}{0pt}\setlength{\rightmargin}{0pt}
\setlength{\leftmargin}{1.2em}\setlength{\parsep}{0pt}}}{
\end{list}}

\newcommand{\nop}[1]{}

\setcopyright{none} 
\begin{document}
\settopmatter{printacmref=false}
\renewcommand\footnotetextcopyrightpermission[1]{}

\title{Privacy for All: Demystify Vulnerability Disparity of Differential Privacy against Membership Inference Attack}

\author{Bo Zhang, Ruotong Yu, Haipei Sun, Yanying Li, Jun Xu and Hui Wang}
\affiliation{%
  \institution{Stevens Institute of Technology}
}

\begin{abstract}
Machine learning algorithms, when applied to sensitive data, pose potential threat to privacy. A
growing body of prior work has demonstrated that membership inference attack (MIA) can disclose specific
private information in the training data to an attacker. Meanwhile, the algorithmic fairness of machine learning has increasingly caught attention from both academia and industry. Algorithmic fairness ensures that the machine learning models do not discriminate a particular demographic group of individuals (e.g., black and female people). Given that MIA is indeed a learning model, it raises a serious concern if MIA ``fairly'' treats all groups of individuals equally. In other words, whether a particular group is more vulnerable against MIA than the other groups.

This paper examines the algorithmic fairness issue in the context of MIA and its defenses. First, for fairness evaluation, it formalizes the notation of {\em vulnerability disparity} (VD) to quantify the difference of MIA treatment on different demographic groups. Second, it evaluates VD on four real-world datasets, and shows that VD indeed exists in these datasets. Third, it examines the impacts of differential privacy, as a defense mechanism of MIA, on VD. The results show that although DP brings significant change on VD, it cannot eliminate VD completely. Therefore,  
fourth, it designs a new mitigation algorithm named \tool\ to  reduce VD. An extensive set of experimental results demonstrate that  \tool\ can effectively reduce VD for both with and without the DP deployment.
\end{abstract}

\maketitle
\section{Introduction}
\label{sc:intro}
Research and advances in the field of machine learning has resulted in algorithms and technologies for improving cybersecurity by helping identifying security threats and system vulnerabilities \cite{seo2018gids,yamaguchi2011vulnerability,she2019neuzz}. However, a line of recent research has shown that  machine learning also can enable novel and sophisticated privacy attacks that leak information about the training dataset \cite{fredrikson2015model,rahman2018membership,song2018natural}. Particularly, one such attack called {\em membership inference attack} (MIA) can infer whether an individual record is included in the model’s training dataset \cite{shokri2017membership} by using machine learning techniques. 
It has been shown that MIA can lead to serious privacy concerns when the training data contains sensitive information (e.g., medical records and financial information). 

The problem of algorithmic fairness of machine learning has attracted considerable attention from both academic and industry in the recent years. The key of algorithmic fairness is to ensure that a  machine learning model  does not discriminate against a particular demographic group (e.g., black people). 
There have been several cases showing that the current ML models are indeed discriminated. For example, as shown by a recent study \cite{dressel2018accuracy}, the current criminal risk assessment tool named COMPAS (standing for Correctional Offender Management Profiling for Alternative Sanctions) predicts the criminals' recidivism risk (i.e., the criminal will re-offend) within 2 years indeed discriminates black defendants - black defendants who did not recidivate were incorrectly predicted to re-offend is nearly twice as high as their white counterparts. 
To address the fairness issue, enormous efforts have been spent on defining fairness models and metrics \cite{calders2009building,feldman2015certifying,biddle2006adverse,feldman2015certifying} as well as developing  fairness-enhancing machine learning algorithms \cite{calders2009building,kamiran2009classifying,kamishima2011fairness,zafar2017fairness}. It has been identified that one of the main reasons of bias in learning is the heavy imbalance of different groups in the training data \cite{dwork2018group}. For example, the underlying cause of the discrimination by COMPAS is the imbalanced distribution between black and white defendants in the training dataset. 

Many real-world datasets that the prior MIA works have explored  not only contain demographic information of individuals but also are heavily imbalanced among different demographic groups (as we will show in the empirical study).  Therefore, MIA, as a learning model, faces the same concern if it is biased towards some particular demographic groups (e.g., black and female people). Therefore, one of the fundamental questions that has to be investigated is whether some demographic groups are more vulnerable to MIA than other groups, due to the skewed distribution of the input data. 

The fairness issue is also related to the defense mechanisms against MIA. In the literature, multiple defense mechanisms against MIA (e.g.,  \cite{nasr2018machine,salem2019ml,jia2019memguard}) have been explored. However, none of these defense mechanisms can provide a theoretical privacy guarantee against MIA. On the other hand, differential privacy (DP) \cite{dwork2011differential} provides a  rigorous privacy guarantee against MIA if the training process is differentially private \cite{shokri2017membership}. 

However, DP only provides a theoretical bound of such privacy protection. Typically, the application of DP does not specifically consider the demographic features of individuals, whereas these features are important for the fairness measurement. Thus it remains unclear how applying DP on the target model will impact the fairness of MIA. 

In summary, in this paper, we address the following important research questions: 
\begin{itemize}
\item Does MIA treat different demographic groups unfairly such that some demographic groups are more vulnerable than the others? If it does, what is the underlying cause of such unfairness?  
\item What is the impact of differential privacy, as a defense mechanism against MIA, on MIA's treatment of different groups? Will it reduce the vulnerability disparity by MIA? 
\item How to design effective bias mitigation methods to reduce the vulnerability disparity, before and after applying DP as the defense? 
\end{itemize}

{\bf Our contributions.} Fairness and privacy are two equally important issues of machine learning. Most of the existing studies have investigated these two issues separately. To our best knowledge, this is the first work that studies algorithmic fairness in the context of membership inference attack and its defenses. We make the following contributions:
\begin{itemize}
\item We formalize the notion of {\em vulnerability disparity} (VD) to quantify the degree of fairness in MIA's prediction results. VD is adapted from the state-of-the-art fairness definition named {\em equal opportunity} \cite{hardt2016equality}. 
Intuitively, VD measures the difference in the success probability of MIA for different demographic groups. 
\item On four real-world datasets, we evaluate VD of MIA for different race and gender groups. Our results prove the existence of VD for both without DP and with DP applied on the target model. The results also show that DP incurs unpredictable, sometimes significant, change of VD. 
\item We analyze the underlying cause of VD as well the impacts of DP on VD under different privacy settings, and demonstrate that VD is largely determined by the data distribution of the training data, for both without and with DP deployment. 
\item We propose a new reweighing method named \tool{} to mitigate VD. The key idea of \tool\ is to reduce the disparity of data distribution among different groups by a user-specified threshold, where the threshold controls the fairness of MIA output. The experiments result show that \tool{} can greatly reduce VD while preserving the utility of the target model. 
\end{itemize}

The paper is organized as following. Section \ref{sc:pre} introduces the preliminaries. Section \ref{sc:dp-agaist-mia} presents the details of implementing DP as the defense mechanism against MIA. Section \ref{sc:understanding-vd} demonstrates the existence of vulnerability disparity and detailed analysis.  Section \ref{sc:mitigation} discusses the details of our mitigation algorithm. Section \ref{sc:related} discusses the related work.  Section \ref{sc:conclusion} concludes the paper. 

\begin{figure}[!tb]
\vspace{-0.1in}
\centering
\includegraphics[width=0.45\textwidth]{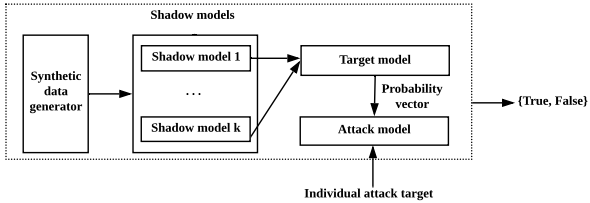}
\vspace{-0.2in}
\caption{\vspace{-0.25in}\label{fig:mia-workflow}Framework of membership inference attack}
\end{figure}

\section{Preliminaries}
\label{sc:pre}
In this section, we elaborate on the preliminaries of membersihip inference attack (MIA), differential privacy (DP), and algorithmic fairness in the literature. 

\subsection{Membership Inference Attack}
\label{subsc:mia}
Membership inference attack (MIA) is a privacy-leakage attack that predicts whether a given record was used in training a target model~\cite{shokri2015privacy}. It works under a setting where the target model is opaque but remotely accessible, and it requires the target model to report the prediction probability of different classes.

Technically, MIA follows a workflow as shown in Figure~\ref{fig:mia-workflow}. Given a target model $\Target$ which can be considered as a black-box to the attacker, the attack starts with synthesizing data that mirrors the training samples of $\Target$. The most general means of data synthesization is to initialize a random sample and gradually improve its quality by using the output of $\Target$. On basis of the synthesized data, MIA creates a group of shadow models to approximate the target model. Each shadow model is trained with the samples that are randomly picked from the synthesized data as well as the output of these samples by $\Target$.  
Aided by the shadow models, MIA derives the final attack model, which  is typically a binary classifier in the format of a neural network. This model is trained by taking as the inputs as the shadow models' prediction probability on the initially synthesized data and the output as considering whether those samples participated in the training of the shadow models. As demonstrated by Shokri \etal~\cite{shokri2015privacy}, the attack models constructed in such an manner can produce high attacking fidelity. As $\Target$ outputs more than 1 label class, MIA indeed consists of multiple attack models, each pertaining to a unique classification label class of $\Target$. 
In summary and formally, MIA can be described as: 
\[\MIA: x^{*}, \mathcal{A}[\Target] \rightarrow \{0, 1\},\]

where $\mathcal{A}[\Target]$ represents the access to the target model $\Target$, $x$ is an arbitrary data point, and 0 means MIA predicts $x^{*}$ as a non-training sample of $\Target$ and 1 otherwise. 

\subsection{Algorithmic Fairness}
\label{sbsec:eq-opp}
The problem of algorithmic fairness in machine learning (ML) algorithms has attracted considerable attention in the recent years.  
Formally, given a dataset consisting of $n$ i.i.d. samples $\{(A_i, X_i, Y_i)\}_{i=1}^n$ from a joint distribution $p_{A,X,Y}$  with domain $A \times X \times Y$, where $A$ denotes one or more protected (discriminatory) features (e.g., gender and race), $X$ denotes other non-protected features used for decision making, and $Y$ denotes the outcome feature. A fair ML system should ensure that the predicted values $\hat{Y}$ of $Y$ do not have discriminatory effects towards particular groups (defined by the associated $A$ values) or individuals. In this paper, we use $A=1$ ($A=0$, resp.) to denote the {\em protected} (unprotected, resp.) group. In the fairness community, {\em Gender=``female''} and {\em Race=``black''} are typically considered as the protected groups on the sensitive attributes {\em Gender} and {\em Race}. 
The ML community has proposed a multitude of formal, mathematical definitions of fairness. 
Intuitively, these fairness definitions are concerned with the protected groups (such as racial or gender groups) and require  some statistic of interest be approximately equalized across different groups. Standard choices for these statistics include positive classification rates \cite{calders2010three}, false positive or false negative rates \cite{hardt2016equality,KleinbergMR17} and positive predictive value \cite{chouldechova2017fair}.

In this paper, we consider a commonly-used fairness measurement named~\emph{equal opportunity} \cite{hardt2016equality}, which requires that protected and unprotected groups should have equal true positive rate. We pick this fairness definition among a number of candidates (e.g., statistical disparity \cite{calders2009building} and equalized odds \cite{hardt2016equality}) because we are only interested in the true positive results by MIA (i.e., the member records that are identified by MIA).  
Formally, {\em equal opportunity} is defined based on {\em disparity} $DI$, which is measured as below:

\begin{equation}
\begin{aligned}
\label{eqn:stadisparity}
   DI = Pr(\hat{Y}=1 |A=1, Y=1)  - Pr(\hat{Y}=1 |A=0, Y=1),
\end{aligned}
\end{equation}

where $A=1$ ($A=0$, resp) denotes the protected (unprotected) group respectively, and $\hat{Y}\in\{0, 1\}$  denotes the predicted value of label $Y$. We only consider the binary classification (i.e., $\hat{Y}\in\{0, 1\}$) as the MIA model is a binary classifier. 
Intuitively,  given the target outcomes, $DI$ quantifies the discrepancy between the true positive rate on the protected group ($A=1$) and the unprotected group ($A=0$) based on their protected characteristic. While $DI=0$, $Y$ satisfies {\em equal opportunity} \cite{hardt2016equality}, indicating enforcement of equal discrepancies across different groups.

\subsection{Differential Privacy}
\label{subsc:pbudget}
Differential privacy (DP) is the de facto standard in measuring the disclosure of privacy pertaining to individuals in a group~\cite{dwork2011differential}. Roughly speaking, a differentially private algorithm ensures that the inclusion or exclusion of an individual record does not significantly affect the result of arbitrary analysis.  Formally, given an arbitrary data domain $X$, and an arbitrary output domain $T$, a randomized algorithm $F$ that maps a dataset to output 
satisfies $\epsilon$-DP if for all dataset pairs $D_1$ and $D_2$ differing on at most one record, and for all subsets $S$ of the output domain $T$: 
\[Pr[F(D_1) \in S] \leq e^{\epsilon} \times Pr[F(D_2) \in S],\]
where the probability is taken over the randomness of $F$, and  $\epsilon$ (called \textit{privacy budget}) specifies the level of guaranteed privacy. Intuitively, larger (smaller, resp.) $\epsilon$ providers weaker (stronger, resp.) privacy protection.

\section{Defending MIA with Differential Privacy}
\label{sc:dp-agaist-mia}

\definecolor{mygray}{gray}{0.9}
\begin{table}[t!]
\centering
\small
\caption{Details of data-sets used in our study} \label{tab:data-des} 
\begin{threeparttable}
\begin{tabular}{l|cccc}
\toprule[0.5pt]
\toprule[0.5pt]
\bf{\emph{Data-set}} & \bf{\# of Record} & \bf{\# of Attr} & \bf{\# of Class} & \bf{Prot Attr}\tnote{1}   \\ \hline
Adult     & 4.5K     & 14    & 2     & G \& R  \\
\rowcolor{mygray}
COMPAS      & 24.5K     & 10    & 2    & G \& R  \\ 
Broward    & 17.2K      & 8    & 2     &  G \& R \\
\rowcolor{mygray}
Hospital    & 110K    & 20   & 4    & G \& R  \\
\bottomrule[0.5pt]
\bottomrule[0.5pt]
\end{tabular}
\begin{tablenotes}
		\item[1] {\bf Prot Attr} indicates the protected attributes, with G and R meaning Gender and Race.
\end{tablenotes}
\end{threeparttable}
\vspace{-2ex}
\end{table}

To serve the objectives of our study, we first reproduce MIA attacks on several real-world datasets and demonstrate the use of DP as a defense. This section covers the details.

\subsection{Reproduction of MIA} 
\label{subsec:miaatt}
In general, the reproduction of MIA follows the workflow as described in Section~\ref{subsc:mia}. In the following, we explain the important specifics. 

\noindent{\bf Data collection and pre-processing.} We use four real-world datasets\footnote{Both Adult and Hospital datasets are used in the original evaluation on MIA~\cite{shokri2015privacy}.} that were widely  used by the fairness and privacy communities. We introduce these four datsets briefly:
\begin{ditemize}
    \item {\em Adult}\footnote{Adult dataset: https://archive.ics.uci.edu/ml/datasets/Adult} dataset that includes 48842 instances and 14 attributess describing information about individuals from the 1994 U.S. census. The prediction task is to determine whether a person makes over 50K a year.
   \item {\em COMPAS}\footnote{COMPAS dataset: https://github.com/propublica/COMPAS-analysis/} dataset contains criminal history, jail and prison time, demographics and COMPAS (which stands for Correctional Offender Management Profiling for Alternative Sanctions) risk scores for defendants from Broward County,  Florida. The prediction task is to infer a criminal defendant's likelihood of becoming a recidivist (i.e., a criminal who re-offend). 
   \item {\em Broward}\footnote{Broward dataset: https://farid.berkeley.edu/downloads/publications/scienceadvances17/} dataset is a refined and cleaned version of COMPAS dataset. It can achieve better MIA accuracy than COMPAS dataset. 
   \item {\em Hospital}\footnote{Hospital dataset: https://www.dshs.texas.gov/THCIC/Hospitals/Download.shtm} dataset stores the hospital discharge data of the inpatients stays in several health facilities and released by the Texas Department of State Health Services from 2006 to 2009. The dataset includes the  demographic information such as the gender, age, race of  the patients, their treatments, and length of stay. The prediction task is to infer the patient’s main procedure based on the attributes other than secondary procedures.
\end{ditemize}
More details of these four datasets are summarized in Table \ref{tab:data-des}. We consider two attributes, namely gender (G) and race (R), as they are law-regulated {\em protected  attributes}~\cite{OurDocum57:online} in the fairness community. 
We also further pre-process the four datasets prior to training and testing; the records with missing values are removed and all the features are converted into numeric values. We keep the categorical features and do not encode them into binary values. 

\noindent{\bf Model building.} To facilitate our explanations of vulnerability disparity, we consider the decision-tree based ID3 model~\cite{friedman2010data} as the target model, due to its simplicity and interpretability. We only consider binary classification in this paper. Our target model can be replaced with other ML models.
For all the datasets, 50\% of the samples are used for training and the remaining for testing. The training of ID3 depends on a hyper-parameter that specifies the tree depth. In our study, we tune this parameter until the testing accuracy becomes comparable to previous research and when over-fitting appears. The reason for requiring over-fitting, as the literature unveils, is to preserve success space of MIA.

Recall that our goal is to deploy DP against MIA and understand the vulnerability disparity. Intuitively, the basis of such a goal is the success of MIA. We, therefore, consider one of the strongest attack where the adversary is fully aware of the target model and has access to part of the original training samples. While training the attack model, we reuse the target model as the shadow model. We also respectively pick 15\% from the target model's training and testing samples as the synthesized data\footnote{For the COMPASs dataset, more than 15\% of original samples are used due to the small size of the entire dataset}. We follow the previous research~\cite{shokri2015privacy} and construct the attack model as a neural  network which consists of one hidden full-connection layer with 300 neurons and a soft-max layer for the output.

\definecolor{mygray}{gray}{0.9}
\begin{table}[t!]
\centering
\small
\caption{Results of reproduction of MIA \label{tab:mia-reproduction}}
\begin{threeparttable}
\begin{tabular}{l|ccccc}
\toprule[0.5pt]
\toprule[0.5pt]
\bf{\emph{Dataset}} & \bf{T-Depth}\tnote{1} & \bf{T-Train}\tnote{2} & \bf{T-Test} & \bf{ MIA Pre}\tnote{3} & \bf{MIA Rec} \tnote{3} \\ \hline
Adult   & 14 & 0.96                 & 0.79                & \bf{$0.627^*$}\tnote{4}     & 0.83     \\
\rowcolor{mygray}
Broward  & 8 & 0.95                & 0.58               & 0.63    & 0.96     \\
COMPAS   & 10 & 0.75                 & 0.70                & 0.56     & 0.76 \\
\rowcolor{mygray}
Hospital & 15 & 0.89                 & 0.56                & 0.636     & 0.67  \\
\bottomrule[0.5pt]
\bottomrule[0.5pt]
\end{tabular}
\begin{tablenotes}
        \item[1] {\bf T-Depth} indicates depth of the decision tree that works as the target model
        \item[2] {\bf T-Train} indicate Target-Train
        \item[3] {\bf Pre} and {\bf Rec} indicate precision and recall of MIA
		\item[4] {\bf*} indicates results outperforming the original evaluation of MIA~\cite{shokri2015privacy}
\end{tablenotes}
\end{threeparttable}
\vspace{-2ex}
\end{table}

\noindent{\bf Evaluation metrics.} Our evaluation metrics include: (1)  the accuracy of the target model on training and testing data respectively, and (2) precision and recall of MIA. 
The accuracy of the target model on the training (testing, resp.) data is measured as the prediction precision (i.e., the fraction of classification results that are correct) on the training (testing, resp.) data. 
We follow the definition of precision and recall in \cite{shokri2017membership} to evaluate the performance of MIA. The precision of MIA is calculated as the fraction of records inferred as
members are indeed members of the training dataset, and the recall is measured as the fraction of the members that are correctly inferred as members by MIA.

\noindent{\bf Results and analysis.} We respectively select 20\% of the training and testing data (which are not overlapping with the sub-set to train the target model) to evaluate performance of MIA. Table~\ref{tab:mia-reproduction} reports the average precision and recall of MIA by five rounds. Overall, MIA demonstrates effective exploitation while the target model achieve satisfactory accuracy. In particular, for the Adult dataset, both MIA and the target model outperform their counterparts in the initial MIA work~\cite{shokri2015privacy}. It is worth noting that we escalate the fidelity of MIA on Adult dataset from nearly random guess (50.3\%) to a meaningful level (62.7\%). On Hospital dataset, the accuracy of the target model as well as the precision and recall of MIA are comparable to the previous study~\cite{shokri2015privacy}. We could not compare the performance on Broward and COMPAS datasets as they were not used in the previous study~\cite{shokri2015privacy}. 

We have to note that the results of our MIA reproduction comply with the common understanding of the community that the feasibility of MIA is attributable to over-fitting of the target model. Considering the gap between training accuracy and testing accuracy as the metric, all our four target models carry a certain level of over-fitting, well matching the effectiveness of MIA. Also it is  observable that larger gaps lead to more severe attacks (see Broward \emph{v.s.} COMPAS).  

\subsection{Defending MIA with DP}
\label{subsec:dp}

\begin{algorithm}
\caption{DP-ID3 \cite{friedman2010data}. Operations that enforce DP are highlighted in boxes with red outline.}
\label{alg:dpid3}
\begin{algorithmic}[1]
\STATE{ {\bf procedure} Buid\_SuLQ\_ID3($\mathcal{T}$, $\mathcal{A}$, $\mathcal{C}$, $d$, $\epsilon$)}
\STATE{ \ \ \ \ {\bf Input:} $\mathcal{T}$ - private dataset, $\mathcal{A}=\{A_i,...A_d\}$ - a set of attributes, $\mathcal{C}$ - class attribute, d - maximal tree depth, $\epsilon$ - differential privacy budget}
\STATE{ \ \ \ \ $m = |\mathcal{A}|$ (i.e. \# of attributes)}
\STATE{ \ \ \ \ {\bf if} $\mathcal{A}=\emptyset$ or d=0 
}
\STATE{ \ \ \ \ \ \ \ \ $\mathcal{T}_{c}$=Partition($\mathcal{T}, \forall c \in \mathcal{C}: r_{\mathcal{C}}=c$)}
\STATE{ \ \ \ \ \ \ \ \ $\forall c \in \mathcal{C}: N_{c}=-1.0$}
\STATE{ \ \ \ \ \ \ \ \ \fcolorbox{red}{white}{\begin{minipage}{20em}$\forall c \in \mathcal{C}: while(N_{c}<0)\ N_{c}= |\mathcal{T}_{c}|+Lap(0,1/\epsilon)$\end{minipage}}}
\STATE{ \ \ \ \ \ \ \ \ $S=\sum_{\forall c\in\mathcal{C}}^{}N_c$}
\STATE{ \ \ \ \ \ \ \ \ $\forall c \in \mathcal{C}: P_c = \frac{N_{c}}{S}$}
\STATE{ \ \ \ \ \ \ \ \ {\bf return} a set of probability $P_c$}
\STATE{ \ \ \ \ {\bf end if}}
\STATE{ \ \ \ \ {\bf for } every attribute $A\in \mathcal{A}$ {\bf do}}
\STATE{ \ \ \ \ \ \ \ \ $\mathcal{T}_j=$Partition($\mathcal{T}, \forall j \in A:r_{A} = j$)}
\STATE{ \ \ \ \ \ \ \ \ $\forall j \in A: \mathcal{T}_{j,c}=$ Partition($\mathcal{T}_j, \forall c \in \mathcal{C}: r_{\mathcal{C}}=c$)}
\STATE{ \ \ \ \ \ \ \ \ \fcolorbox{red}{white}{\begin{minipage}{12em}$N_j^A = |\mathcal{T}_j|+Lap(0,2m/\epsilon)$
\end{minipage}}}
\STATE{ \ \ \ \ \ \ \ \ \fcolorbox{red}{white}{\begin{minipage}{12em}$N_{j,c}^A = |\mathcal{T}_{j,c}|+Lap(0,2m/\epsilon)$\end{minipage}}}
\STATE{ \ \ \ \ \ \ \ \ $\overline{V_{A}}=\sum_{j=1}^{|A|}\sum_{c=1}^{|\mathcal{C}|} N_{j,c}^A \cdot log \frac{N_{j,c}^{A}}{N_j^A}$}
\STATE{ \ \ \ \ {\bf end for}}
\STATE{ \ \ \ \ $\overline{A}=argmax_{A}\overline{V_{A}}$}
\STATE{ \ \ \ \ $\mathcal{T}_i = Partition(\mathcal{T}, \forall i \in \overline{A}: r_{\overline{A}}=i)$}
\STATE{ \ \ \ \ $\forall i \in \overline{A}:$ $SubTree_i$ = Build\_SuLQ\_ID3($\mathcal{T}_i, \mathcal{A} \backslash \overline{A}, \mathcal{C}, d-1, \epsilon$)}
\STATE{ \ \ \ \ {\bf return} a tree with a root node labeled $\overline{A}$ and edges labeled 1 to $|\overline{A}|$ each going to $SubTree_i$}
\STATE{{\bf end procedure}}
\end{algorithmic}
\end{algorithm}

\begin{figure*}[!t]
\vspace{-0.1in}
\centering
        \begin{tabular}{@{}c@{}c@{}c@{}c@{}c@{}}
          \includegraphics[width=0.25\textwidth]{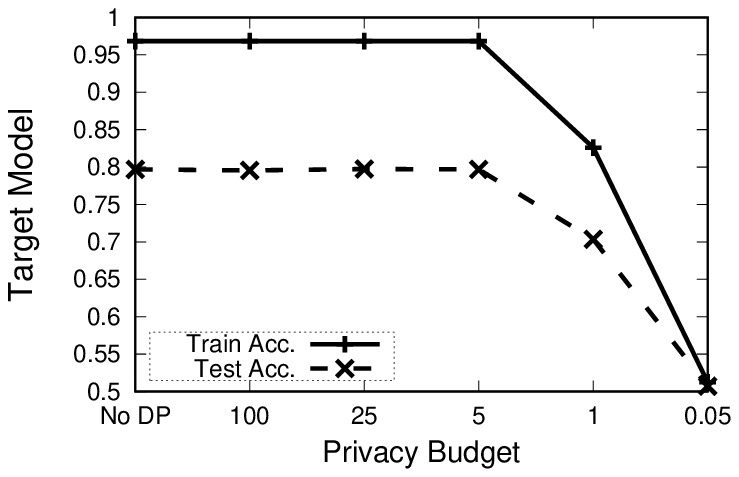}
           &
          \includegraphics[width=0.25\textwidth]{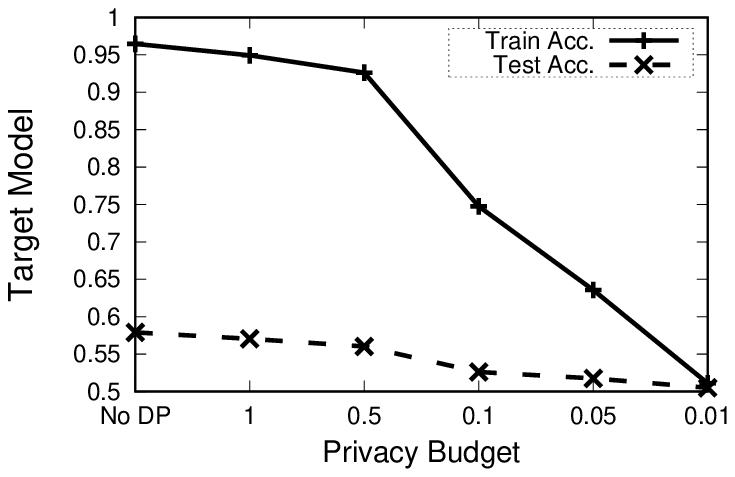}
          &
          \includegraphics[width=0.25\textwidth]{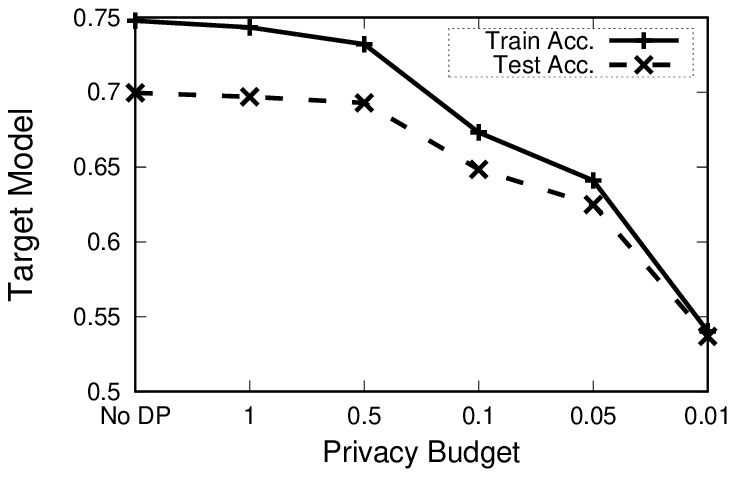}
          &
          \includegraphics[width=0.25\textwidth]{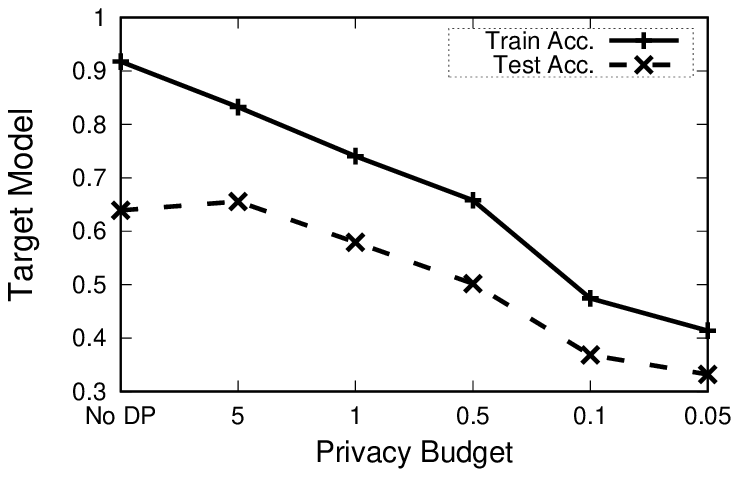}
        \\
        {\scriptsize (a) Adult Dataset (Tree depth=14)}
        &
        {\scriptsize (b) Broward Dataset (Tree depth=11)}
         &
        {\scriptsize (c) COMPAS Dataset (Tree depth=8)}
         &
        {\scriptsize (d) Hospital Dataset (Tree depth=15)}
        \\
        \includegraphics[width=0.25\textwidth]{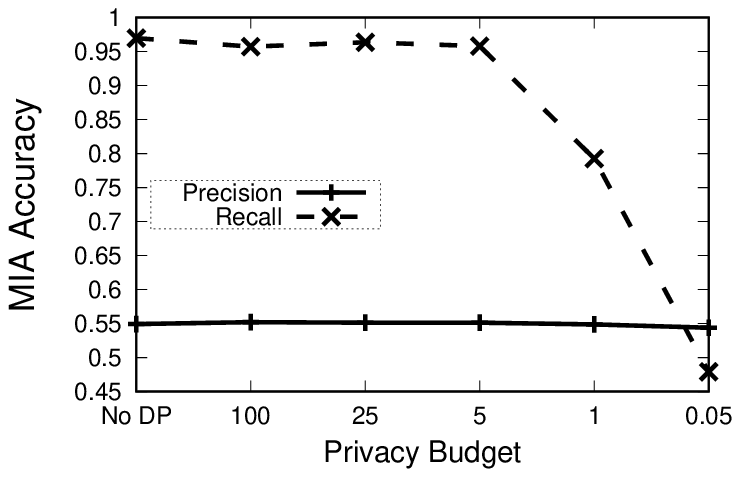}
         &
          \includegraphics[width=0.25\textwidth]{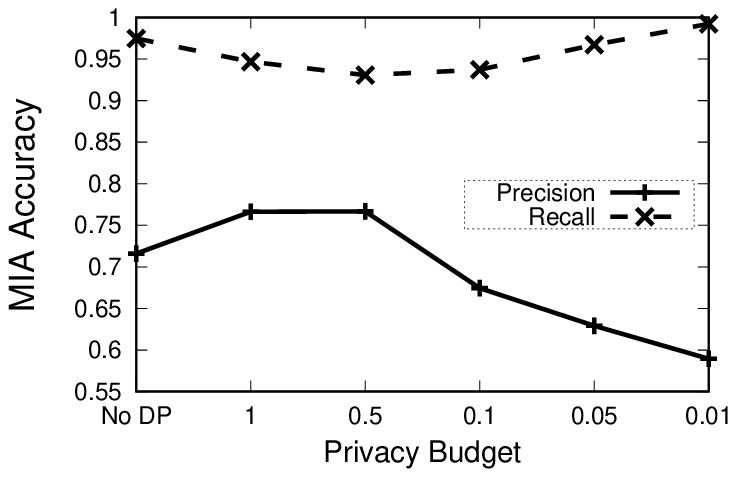}
           &
          \includegraphics[width=0.25\textwidth]{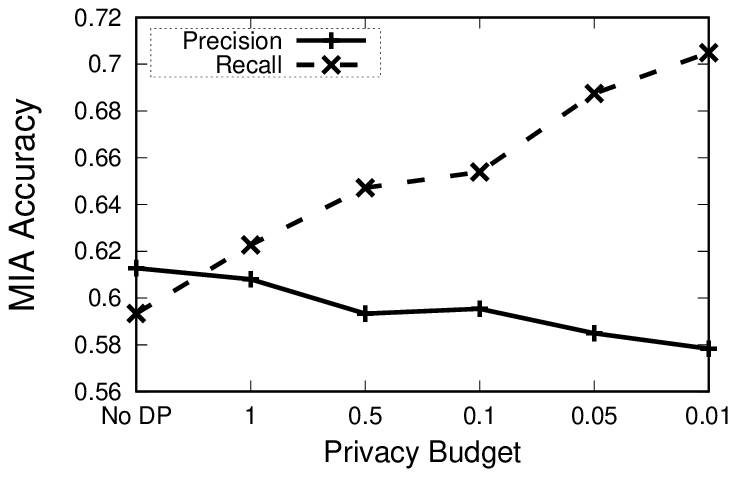}
          &
          \includegraphics[width=0.25\textwidth]{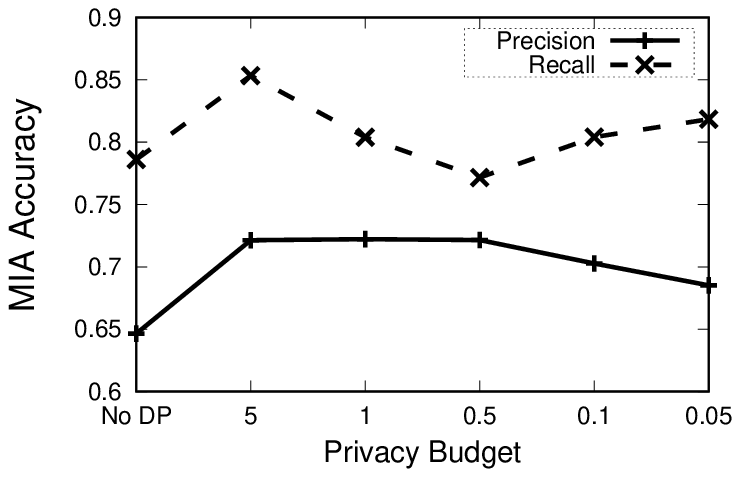}
        \\
        {\scriptsize (e) Adult Dataset (Tree depth=14)}
        &
        {\scriptsize (f) Broward Dataset (Tree depth=8)}
         &
        {\scriptsize (g) COMPAS Dataset (Tree depth=10)}
         &
        {\scriptsize (h) Hospital Dataset (Tree depth=15)}
        \end{tabular}
        \vspace{-0.2in}
       \caption{\small \label{fig:target-acc} Performance of target model and MIA under different privacy budgets.}
\end{figure*}

\noindent{\bf Design of defense.} As MIA runs solely on the outputs of the target model, making the target model differentially private, by design, can defend against the MIA attack. There are two approaches to enforce DP on the target model: (1) introduce noise to the input data~\cite{chen2015differentially,dwork2009differential} and (2) add perturbation to the target model \cite{jagannathan2009practical,friedman2010data}. The first approach, however, often perturbs the data points, and thus possibly ruins the mapping between the original samples and the differentially private ones. This, as we will explain later, prevents our measurement of vulnerability disparity. Hence, we decide to enforce the ID3 target model with DP by following the design and implementation in \cite{friedman2010data,dwork2006calibrating}. The algorithm is summarized in Algorithm~\ref{alg:dpid3} and we name it {\em DP-ID3}. Briefly speaking, DP-ID3 adds Laplace noise to the count that is used to compute information gain (line {\tt 15-16}) and the voting on the leaf node (line {\tt 7}). DP-ID3  has been proven to satisfy DP~\cite{blum2005practical}. 

The performance of DP-ID3 depends on the setup of  \emph{privacy budget} $\epsilon$. We base the selection of privacy budget on performance of MIA. To be specific, we pick a range of privacy budgets where MIA shifts from being slightly weakened to being nearly disabled. The final ranges of privacy budgets are observable in Figure~\ref{fig:target-acc}; the privacy budget ranges in [0.01, 100].

\noindent{\bf Results and analysis.} In Figure~\ref{fig:target-acc}, we present the precision and recall of the target model and MIA for various privacy budgets. For the sake of normalizing randomness of DP, all the tests are repeated 25 times and we report the average results. Unsurprisingly, both precision and recall of the target model decreases when DP strength increases. However, DP also demonstrates high defense effectiveness against MIA. 
For all the four datasets, as DP strength increases (\emph{i.e.}, the privacy budget decreases), the precision of MIA persistently drop and finally approach the level of random guess. 
We analyze why DP can defend against MIA. In essence, the defense of DP derives from its effects on reducing the over-fitting property of the target model. More specifically, DP introduces Laplace noise to the training process of the target model, which, on one hand, fuzzifies the choice of attribute for tree splitting (line {\tt 15-16} of Algorithm~\ref{alg:dpid3}) and, on the other hand, dilutes the purity of the leaf nodes (line {\tt 7} of Algorithm~\ref{alg:dpid3}). Therefore, DP generalizes the decision boundary of the target model, making it less fit the training data. 

We also notice that on COMPAS dataset, the recall of MIA surprisingly increases quickly while DP gets stronger. We investigated the underlying cause of this phenomenon. It turned out that MIA labels more samples in the testing data as positive when DP gets stronger. Therefore, the recall increases sharply while the precision drops to approximately 0.5 (i.e., random guess).

\section{Understanding of Vulnerability Disparity}
\label{sc:understanding-vd}
In principle, DP only constrains the upperbound of privacy that each individual can lose. However, DP does not ensure that different population groups experience the same amount of actual privacy leakage. In particular,  the imbalanced data distribution of different demographic groups (e.g., white vs black people) may lead to inequitable privacy protection. In this section, we unveil this type of inequitability in using DP as a defense against MIA. 

\vspace{-0.05in}
\subsection{Definition of Vulnerability Disparity}
\label{subsec:DV}
To measure the vulnerability a particular group is against MIA attack, We define the concept of \emph{vulnerability disparity} . Vulnerability disparity (VD) is adapted from the standard fairness definition of {\em equal opportunity}  (Equation~\ref{eqn:stadisparity}). Formally, given a dataset $D$ and a protected attribute $A$, where $A=a$ and $A=\bar{a}$ define the protected and unprotected groups respectively, 

We define the {\em vulnerability disparity} $VD$ of MIA on $D$ as: \begin{equation}
\begin{aligned}
\label{eqn:disc}
VD = P(\hat{Y} = 1|Y=1, A=a) - P(\hat{Y} = 1|Y=1, A=\bar{a}),
\end{aligned}
\end{equation}
where $\hat{Y}$ denotes the binary output label of MIA. In particular, $Y=1$ denotes that the data point is a member of $D$. Intuitively, $VD$ describes the difference of the success probability of MIA for the protected group ($A=a$) versus the unprotected group ($A=\bar{a}$). 
 When $VD>0$ ($VD<0$, resp.), the protected group ($A=a$) is more vulnerable (less vulnerable, resp.) than the unprotected group ($A=\bar{a}$) against MIA. 
 
The definition of VD can be applied to measure the protection of DP against MIA too. Formally, the vulnerability disparity $VD_\mathcal{DP}$ of a differentially private mechanism $\mathcal{DP}$ can be measured by simply adapting Formula \ref{eqn:disc} as following: 
\begin{equation}
\begin{aligned}
\label{eqn:dpdisc}
    VD_\mathcal{DP}=P(\hat{Y}_{\mathcal{DP}} = 1| Y = 1, A=a) 
    - P(\hat{Y}_{\mathcal{DP}}  = 1| Y=1, A=\bar{a}), 
\end{aligned}
\end{equation}
where $\hat{Y}_{\mathcal{DP}}$ denotes the prediction of MIA by using the output of the differentially private target model as input.  
In the ideal case, if a differentially private ML algorithm provides fair group protection, then $VD_\mathcal{DP}=0$. 

To measure the impact of DP on VD, we measure the change of $VD$ before and after the deployment of DP. Formally, let $VD$ and $VD_{\mathcal{DP}}$ be the VD before and after DP deployment on the target model. The {\em change} $c$ of vulnerability disparity is measured as: 
\begin{equation}
\begin{aligned}
\label{eqn:reduction}
   c  = \frac{VD_\mathcal{DP} - VD}{VD},
\end{aligned}
\end{equation}
Intuitively, $c<0$ ($c>0$, resp.) means that the DP mechanism decreases (increases, resp.) VD.

\begin{figure*}[!t]
\vspace{-0.1in}
\centering
        \begin{tabular}{@{}c@{}c@{}c@{}c@{}c@{}}
          \includegraphics[width=0.25\textwidth]{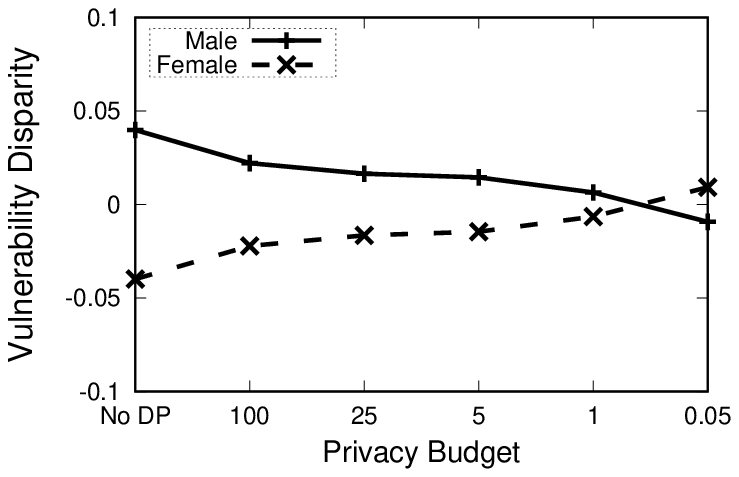}
           &
          \includegraphics[width=0.25\textwidth]{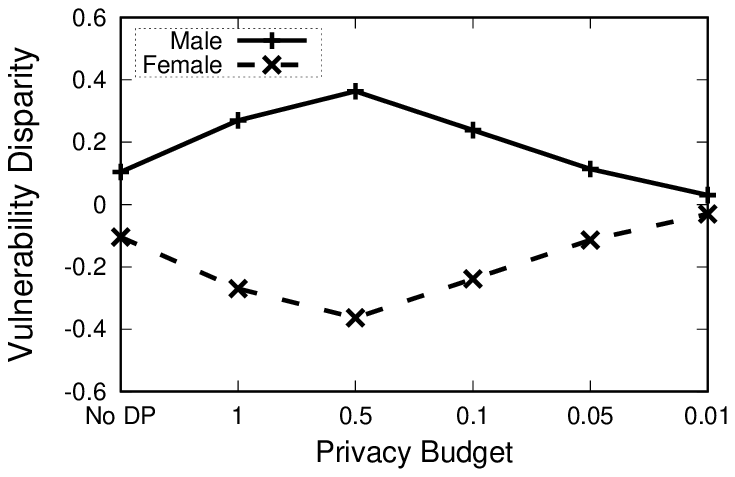}
          &
          \includegraphics[width=0.25\textwidth]{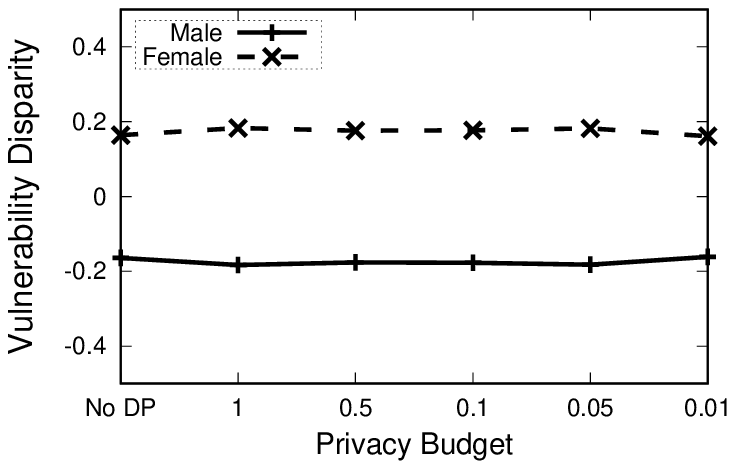}
          &
          \includegraphics[width=0.25\textwidth]{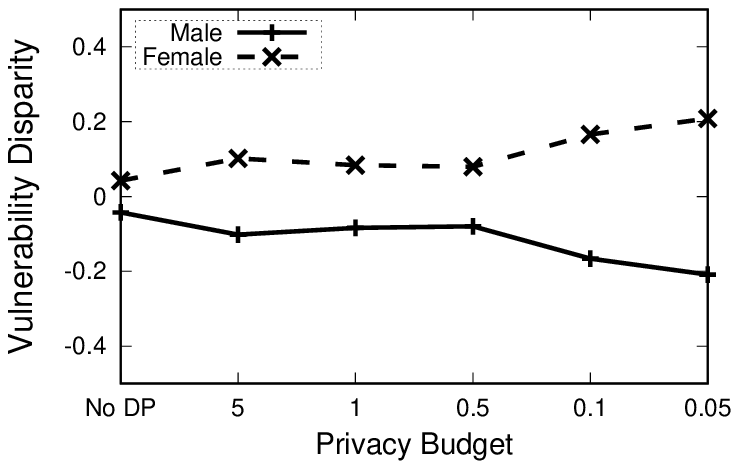}
        \\
       {\scriptsize (a) {\bf ADULT} (Gender) }
        &
        {\scriptsize (b) {\bf BROWARD}  (Gender)}
         &
        {\scriptsize (c) {\bf COMPAS}  (Gender) }
         &
        {\scriptsize (d) {\bf Hospital} (Gender)}
        \end{tabular}
         \\
         \begin{tabular}{@{}c@{}c@{}c@{}c@{}c@{}}
          \includegraphics[width=0.25\textwidth]{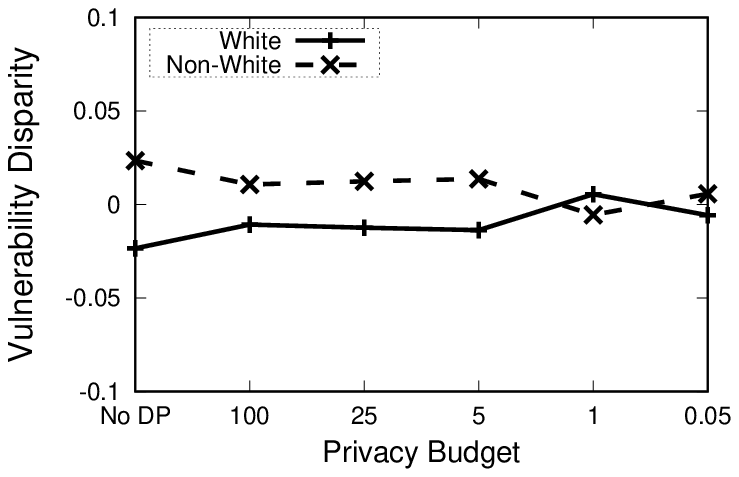}
           &
          \includegraphics[width=0.25\textwidth]{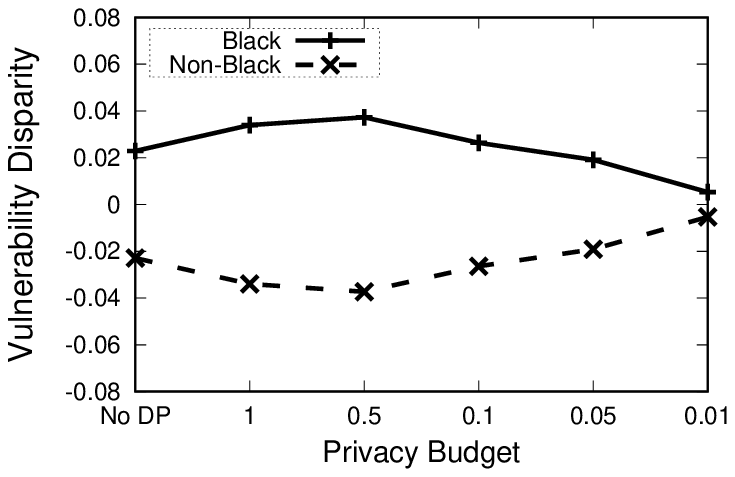}
          &
          \includegraphics[width=0.25\textwidth]{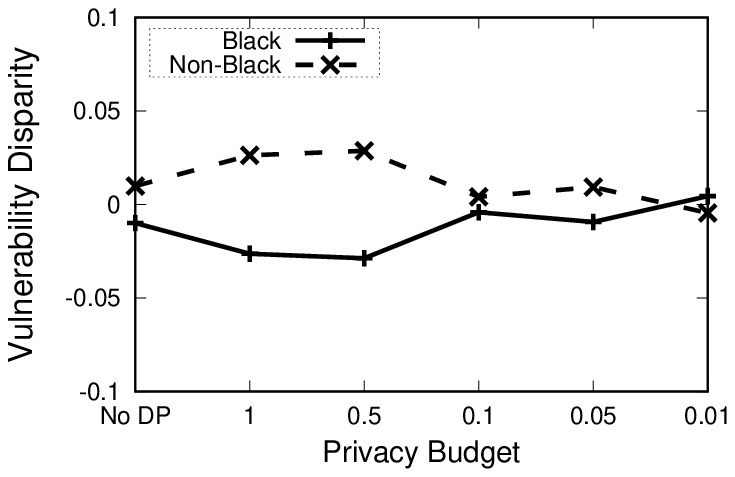}
          &
          \includegraphics[width=0.25\textwidth]{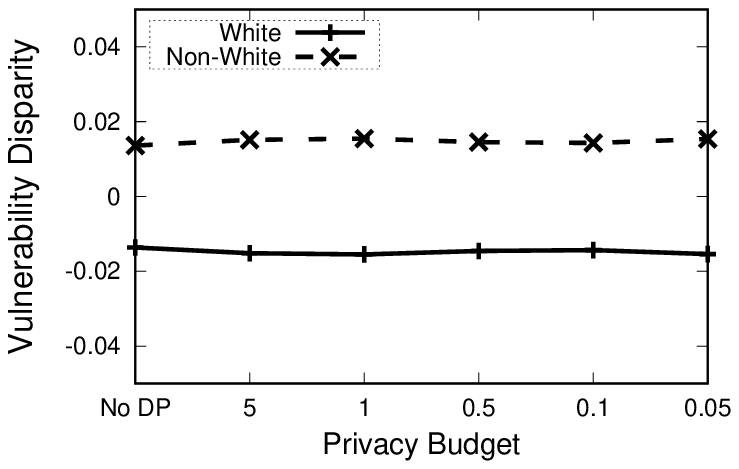}
        \\
        {\scriptsize (e) {\bf ADULT} (Race)}
        &
        {\scriptsize (f) {\bf BROWARD}  (Race)}
         &
        {\scriptsize (g) {\bf COMPAS}  (Race)}
         &
        {\scriptsize (h) {\bf Hospital} (Race)}
        \end{tabular}
        \vspace{-0.1in}
       \caption{\small \label{fig:vul-disp} Vulnerability disparity before and after DP enforcement (``No DP'' means before DP is deployed).}
\end{figure*}

\begin{figure*}[!t]
\vspace{-0.1in}
\centering
        \begin{tabular}{@{}c@{}c@{}c@{}c@{}c@{}}
          \includegraphics[width=0.25\textwidth]{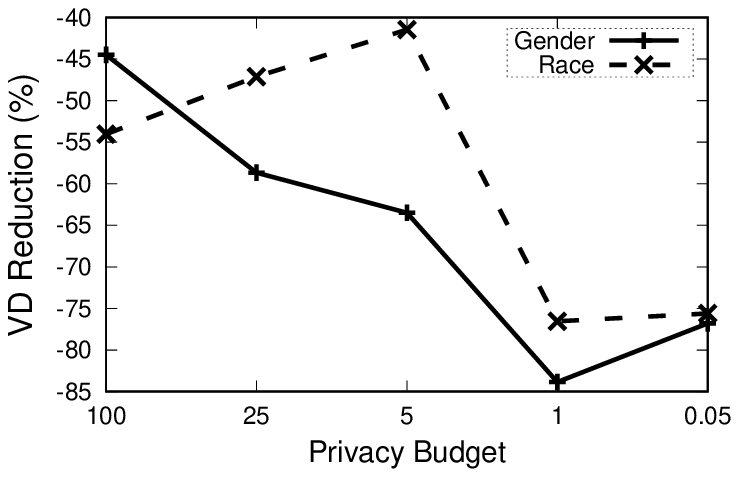}
           &
          \includegraphics[width=0.25\textwidth]{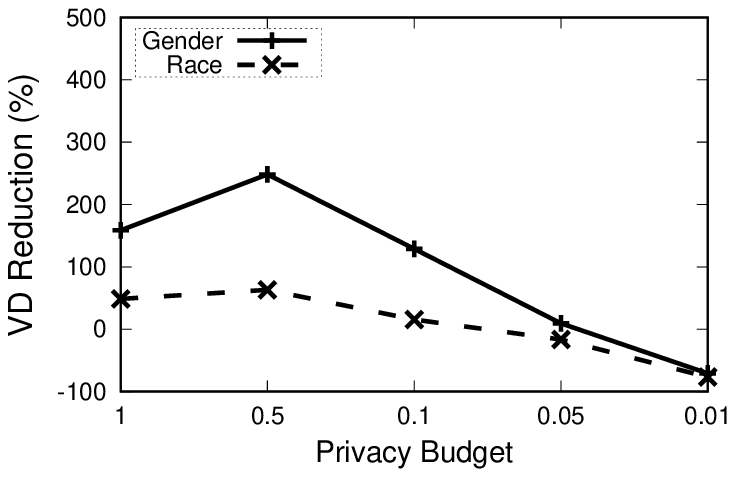}
          &
          \includegraphics[width=0.25\textwidth]{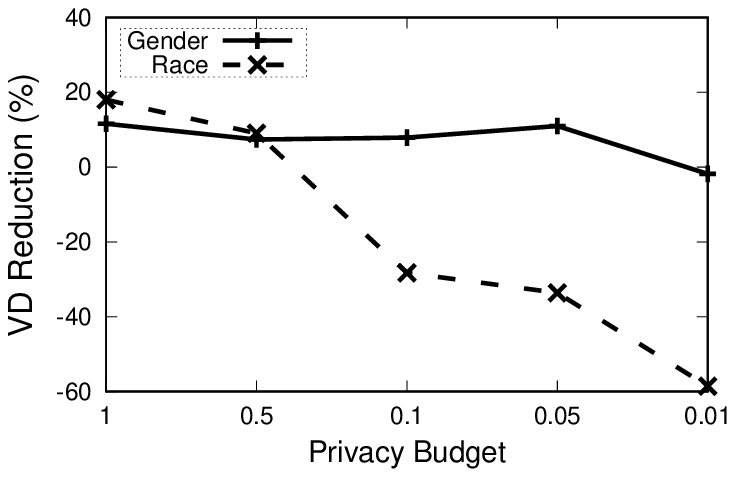}
          &
          \includegraphics[width=0.25\textwidth]{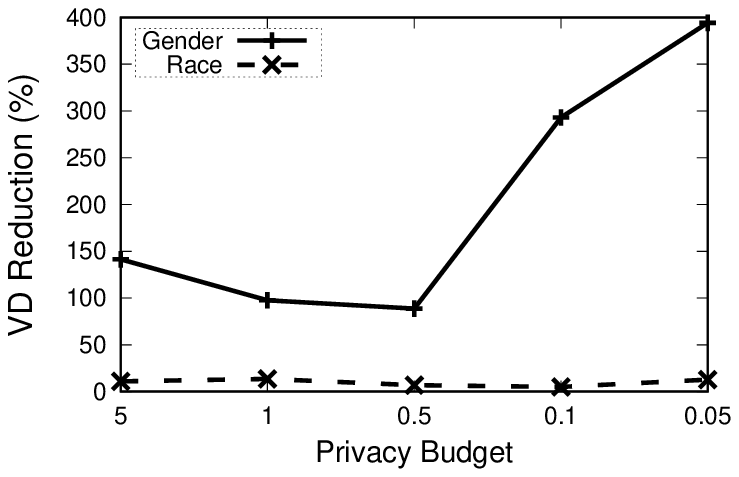}
        \\
       {\scriptsize (a) {\bf ADULT}}
        &
        {\scriptsize (b) {\bf BROWARD}}
         &
        {\scriptsize (c) {\bf COMPAS}}
         &
        {\scriptsize (d) {\bf Hospital}}
        \end{tabular}
        \vspace{-0.1in}
       \caption{\small \label{fig:vul-disp-against-dp} Change of Vulnerability disparity by DP.}
\end{figure*}

\subsection{Measurement of Vulnerability Disparity}
\label{sc:vdmeasurement}
\begin{table}[t!]
\centering
\small
\caption{\label{table:groupdis}Skewness of group distribution}
\vspace{-0.05in}
\begin{tabular}{cc|cc|cc|cc}
\toprule[0.5pt]
\toprule[0.5pt]
 \multicolumn{2}{c|}{\bf Adult} & \multicolumn{2}{c|}{\bf Broward}   & \multicolumn{2}{c|}{\bf COMPAS}  & \multicolumn{2}{c}{\bf Hospital}\\ \hline
Race&	Count&  Race&	Count& Race&	Count&	Race& Count\\
\rowcolor{mygray}
Asian & 4.22k & Hispa. & 1.93k & Hispa.&	3.47k &	Asian & 13.73k \\ 
\rowcolor{white}
Black  &  0.35k & Black & 9.54k & Black & 10.80k & Black & 33.77k \\
\rowcolor{mygray}
White & 38.90k & White& 5.36 k & White & 8.82k & White & 56.66k\\ 
\rowcolor{white} 
Other &  1.73k & Other & 0.41k & Other &  14.46k & Other & 6.06k \\
\toprule[0.5pt]
\toprule[0.5pt]
\end{tabular}\\
(a) {\bf Race Group}\\
\begin{tabular}{cc|cc|cc|cc}
\toprule[0.5pt]
\toprule[0.5pt]
 \multicolumn{2}{c|}{\bf Adult} & \multicolumn{2}{c|}{\bf Broward}   & \multicolumn{2}{c|}{\bf COMPAS}  & \multicolumn{2}{c}{\bf Hospital}\\ \hline
Gen & Count&  Gen&	Count& Gen&	Count&	Gen & Count\\
\rowcolor{mygray}
M &  30.51k & M & 14.71k & M&	16.50k&	M & 76.91k\\ 
\rowcolor{white}
F  &  14.69k & F & 2.54k & F & 8.03k & F & 33.31k\\
\toprule[0.5pt]
\toprule[0.5pt]
\end{tabular}\\
(b) {\bf Gender Group}
\vspace{-0.15in}
\end{table}

In this section, we present the results and  analysis of vulnerability disparity. 
To eliminate the impact of randomness by DP, we ran the experiments 25 times on all four datasets and took the average.

\noindent{\bf Protected attributes and groups.} 
By following the literature on algorithmic fairness research, we pick \emph{race} and \emph{gender} as the protected attributes.  
We must note that in all the four datasets that we use, the classification labels do not have strong correlation with  \emph{race} and \emph{gender}. In other words, the classification results do not strongly depend on these two protected attributes. 
Two groups exist on the gender attribute: male and female. 
The data on the race attribute consists of four groups. However, 
for simplicity, we categorize the data into two race groups only: black and non-black. The non-black group consists of all members that are not black (e.g., White, Asian, and Hispanic). 

\noindent{\bf Skewness of group distribution.}
The distribution of different population groups on the protected attributes of the four datasets are shown in Table \ref{table:groupdis}. Obviously, the group distribution of all the four datasets is heavily imbalanced. Furthermore, the same race group does not always have the same distribution in different datasets. For example, the white group dominates in Adult dataset, while the black group dominates in Broward dataset. 
On gender attribute, the male group always dominates the female group in all the four datasets.
Although it seemly has no direct connection between the vulnerability disparity and race/gender, we will explore if there is any hidden disparity due to the imbalanced data distribution of these two protected attributes.

\noindent{\bf Main results.} The results of VD before and after applying DP on the target model are illustrated in Figure~\ref{fig:vul-disp}. The results of VD reduction for various privacy budgets is shown in Figure \ref{fig:vul-disp-against-dp}. We briefly summarize the main observations below:
\vspace{-0.1in}
\begin{enumerate}[leftmargin=*]
    \item Different datasets show different amounts of VD, for both before and after DP deployment (as shown in Figure \ref{fig:vul-disp}). 
     \begin{ditemize}
         \item {\em Adult} dataset: insignificant VD ($<0.04$) on both race and gender groups before and after DP enforcement. 
         \item {\em Broward} and {\em COMPAS} datasets: significant VD ($\geq 0.1$) on the gender groups before and after DP deployment. Insignificant VD ($\leq 0.04$) on the race groups before and after DP enforcement.
         \item {\em Hospital} dataset: insignificant VD ($<-0.01$) on both gender and race groups before DP. Significant VD ($\geq 0.1$) after DP on gender groups but insignificant VD on race groups after DP enforcement. 
     \end{ditemize}
   \item DP cannot eliminate VD completely on both gender and race attributes. There is no consistent pattern of how VD changes by  DP deployment (as shown in Figure \ref{fig:vul-disp-against-dp}).
   \begin{ditemize}
         \item {\em Adult} dataset:  On both gender and race attributes, VD decreases until the privacy budget decreases to 1. Then VD switches to opposite direction and increases.
         \item {\em Broward} dataset: VD increases after DP deployment on both gender and race groups. The increment is small and stable on race groups, but significant and unsteady on gender groups. 
         \item {\em COMPAS} dataset: VD increases after DP deployment, except when privacy budget is 1 and 0.5. The  pattern is opposite to  {\em Broward} dataset - VD increment is small  and stable on gender groups, but significant and volatile on race groups. 
         \item {\em Hospital} dataset: On gender attributes VD increases to be significant ($>0.1)$ after DP deployment. On race attribute, VD keeps stable. 
     \end{ditemize}
\end{enumerate}

\begin{figure*}[!t]
\vspace{-0.1in}
\centering
        \begin{tabular}{@{}c@{}c@{}c@{}c@{}c@{}}
         \includegraphics[width=0.2\textwidth]{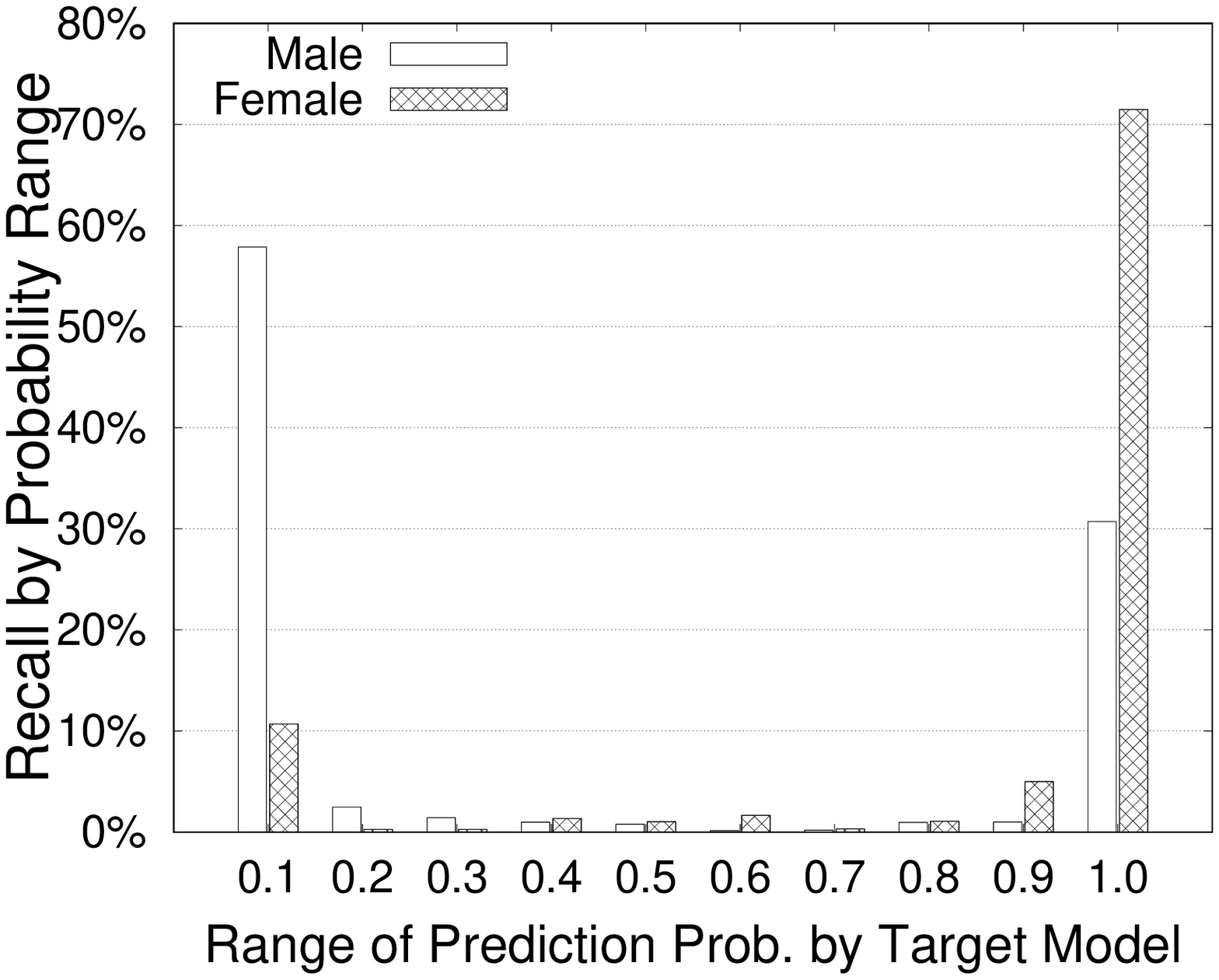}
          &
          \includegraphics[width=0.2\textwidth]{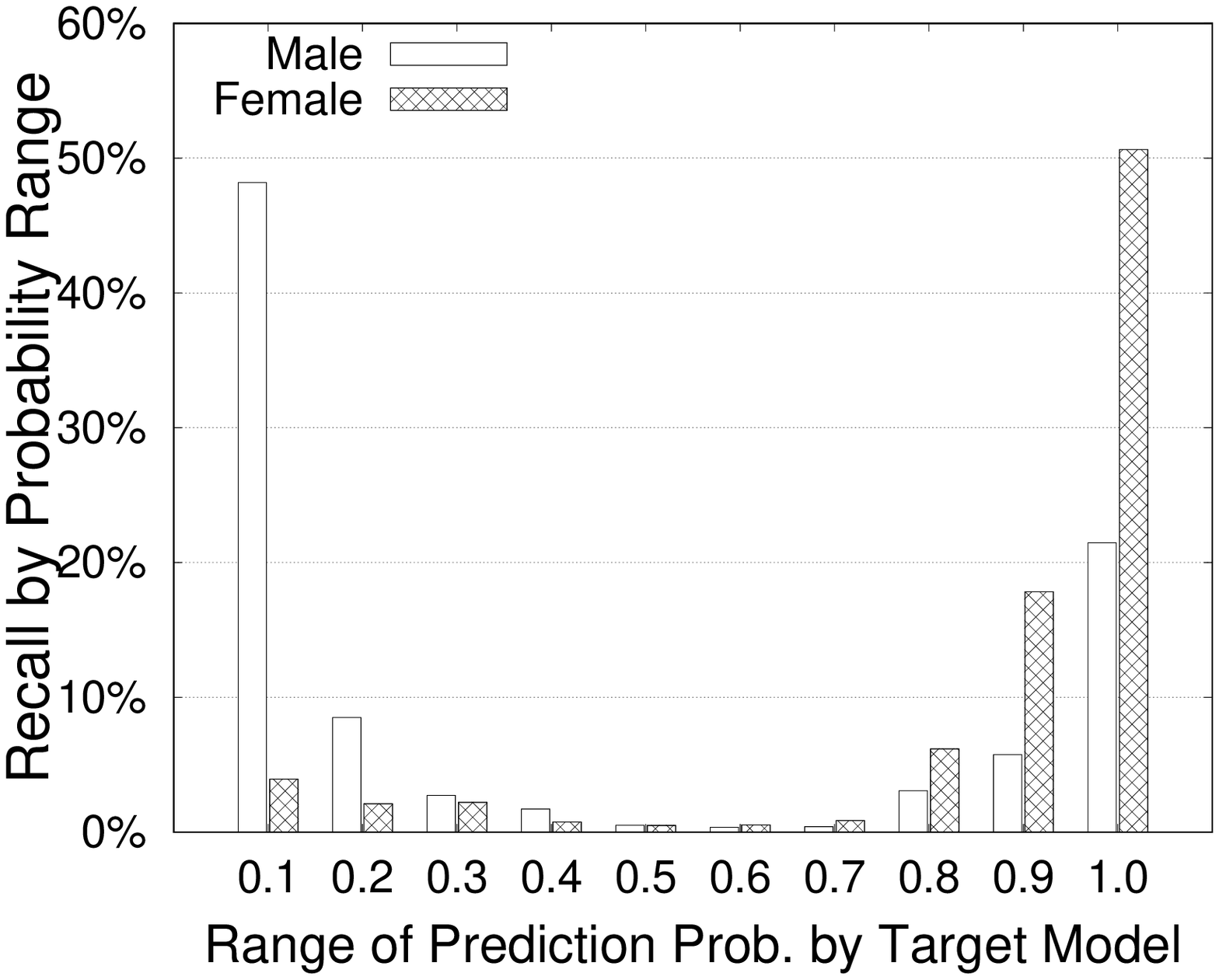}
           &
          \includegraphics[width=0.2\textwidth]{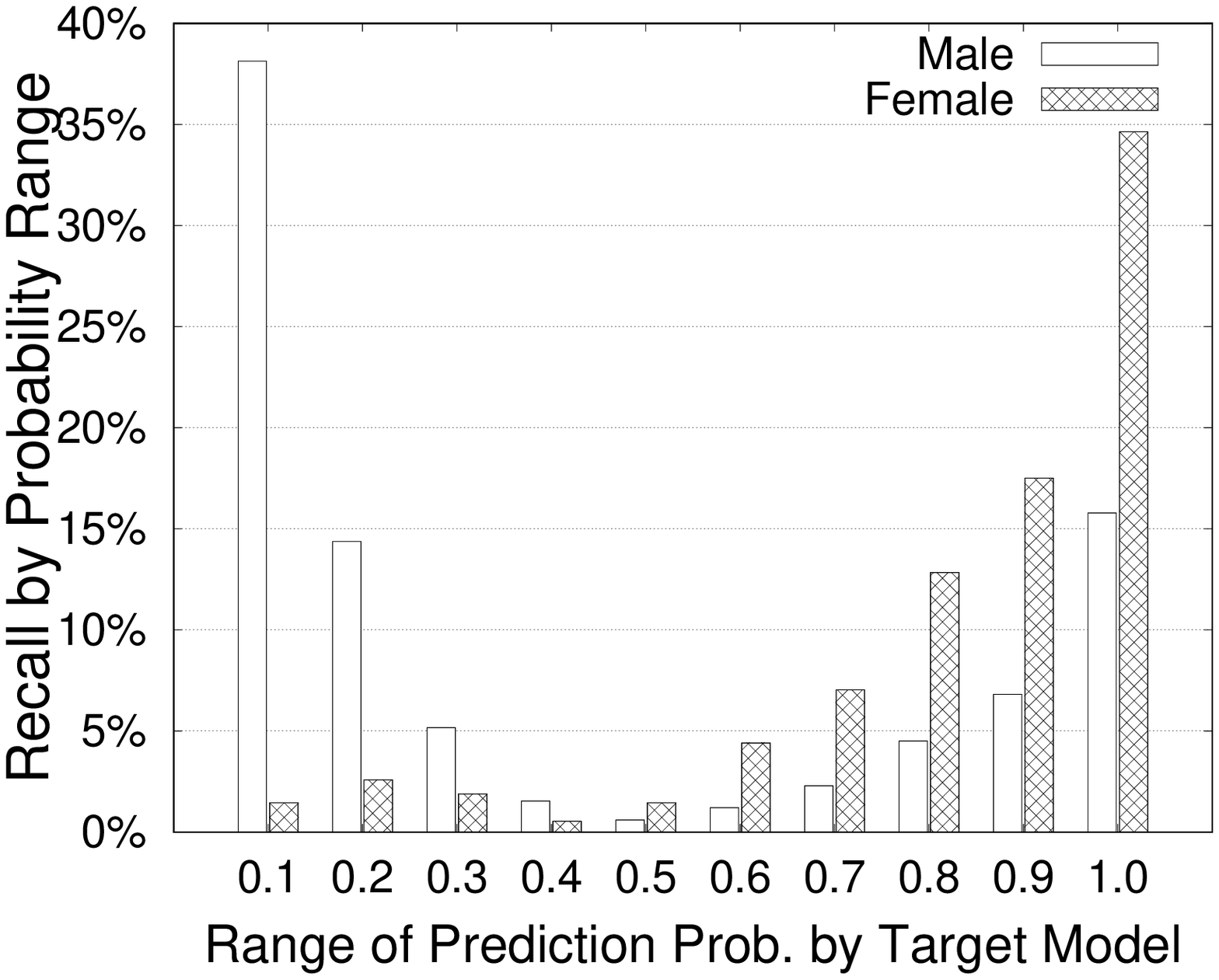}
          &
          \includegraphics[width=0.2\textwidth]{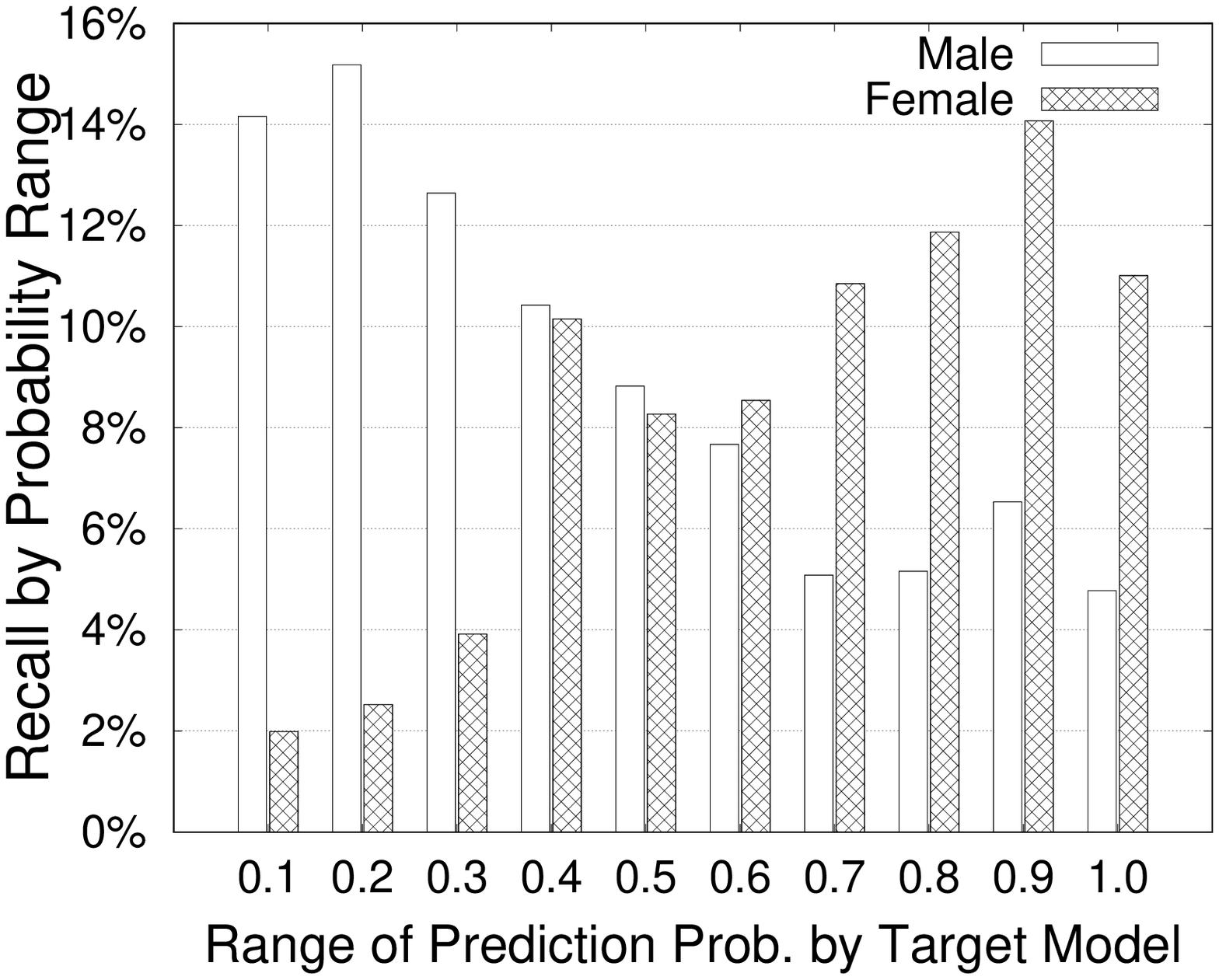}
          &
          \includegraphics[width=0.2\textwidth]{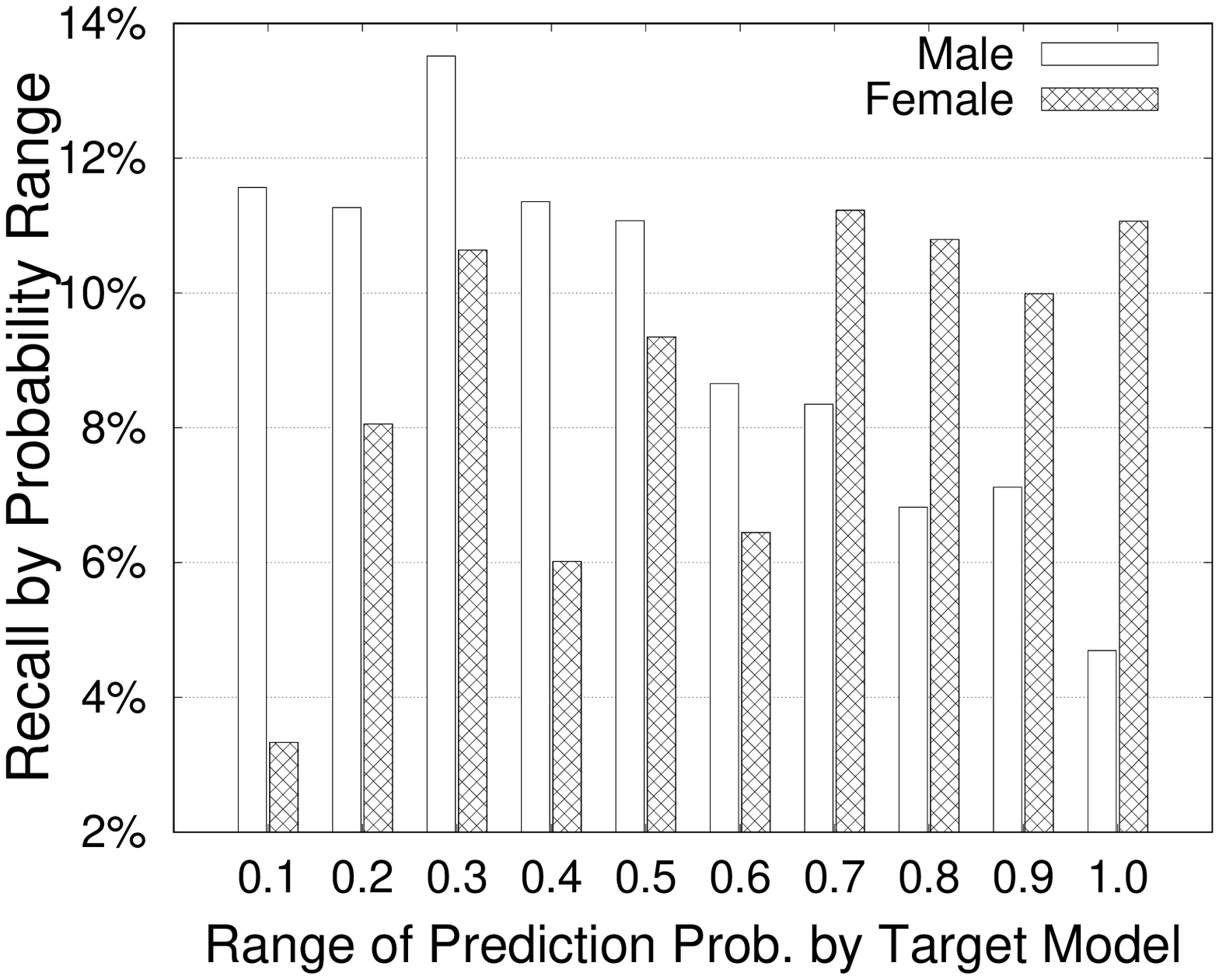}
          \\
       {\scriptsize (a) BROWARD (gender, No DP)}
        &
        {\scriptsize (b) BROWARD (gender, $\epsilon=1$)}
         &
        {\scriptsize (c) BROWARD (gender, $\epsilon=0.5$)}
         &
        {\scriptsize (d) BROWARD (gender, $\epsilon=0.1$)}
         &
        {\scriptsize (e) BROWARD (gender, $\epsilon=0.05$)}
    \\
          \includegraphics[width=0.2\textwidth]{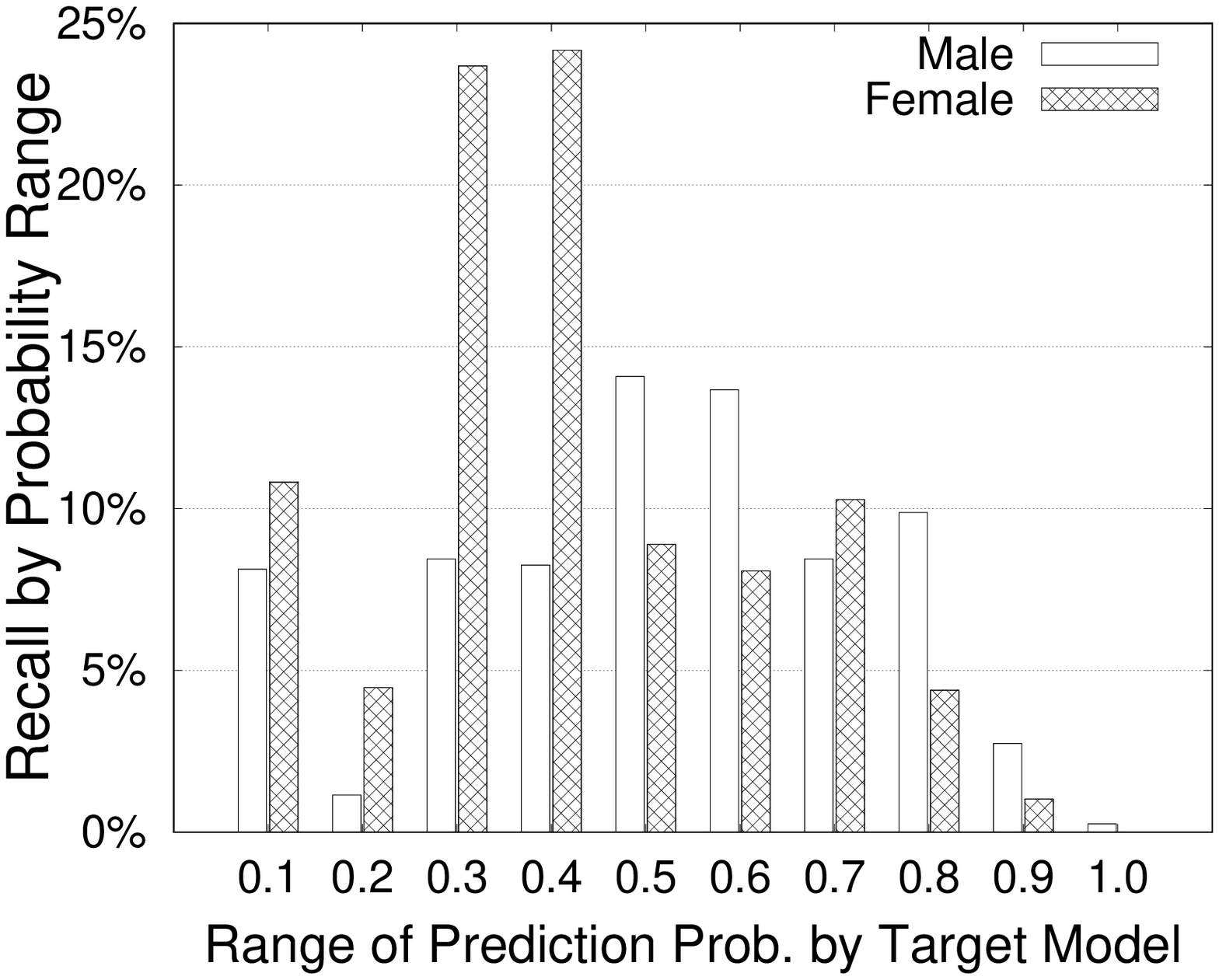}
          &
          \includegraphics[width=0.2\textwidth]{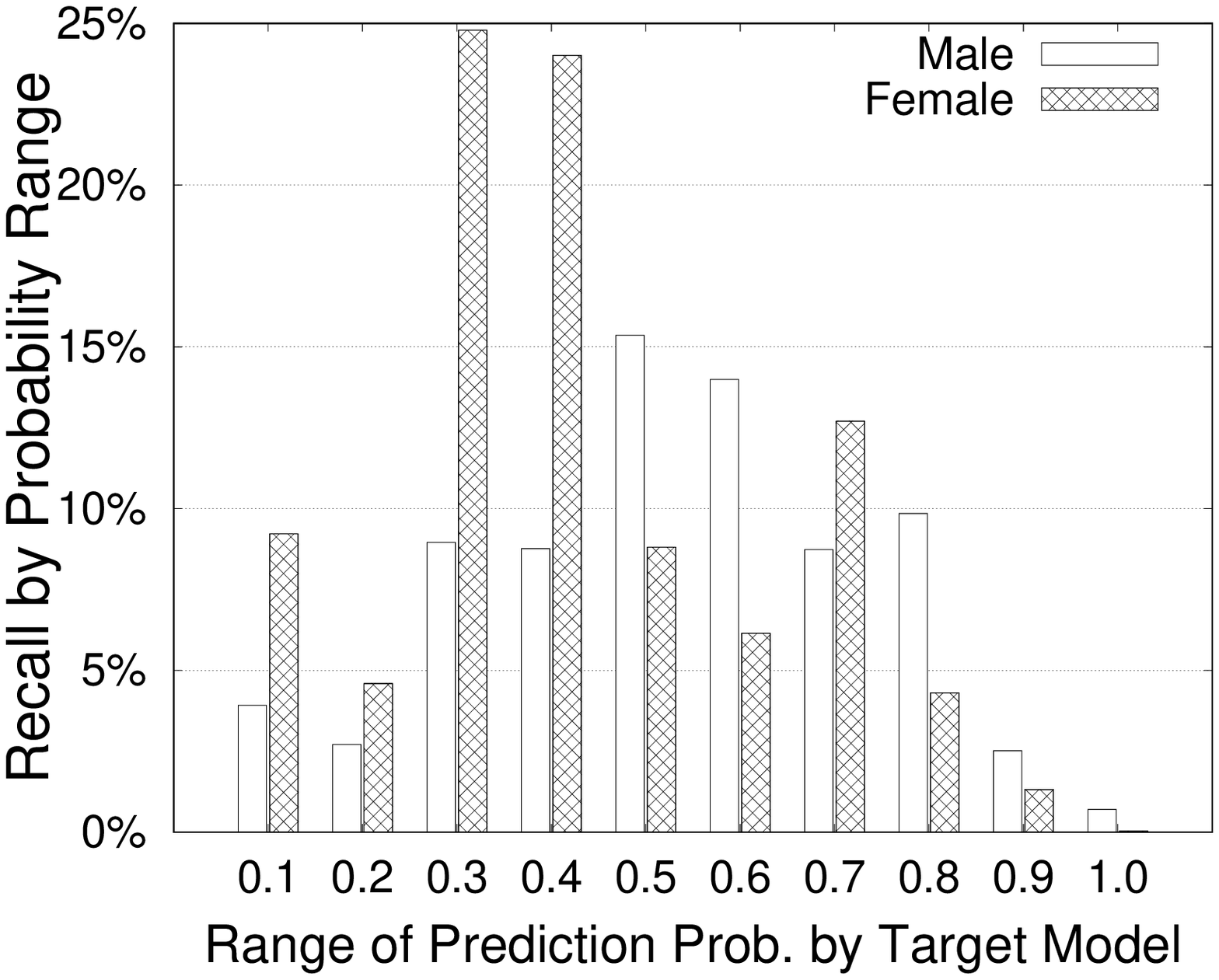}
           &
          \includegraphics[width=0.2\textwidth]{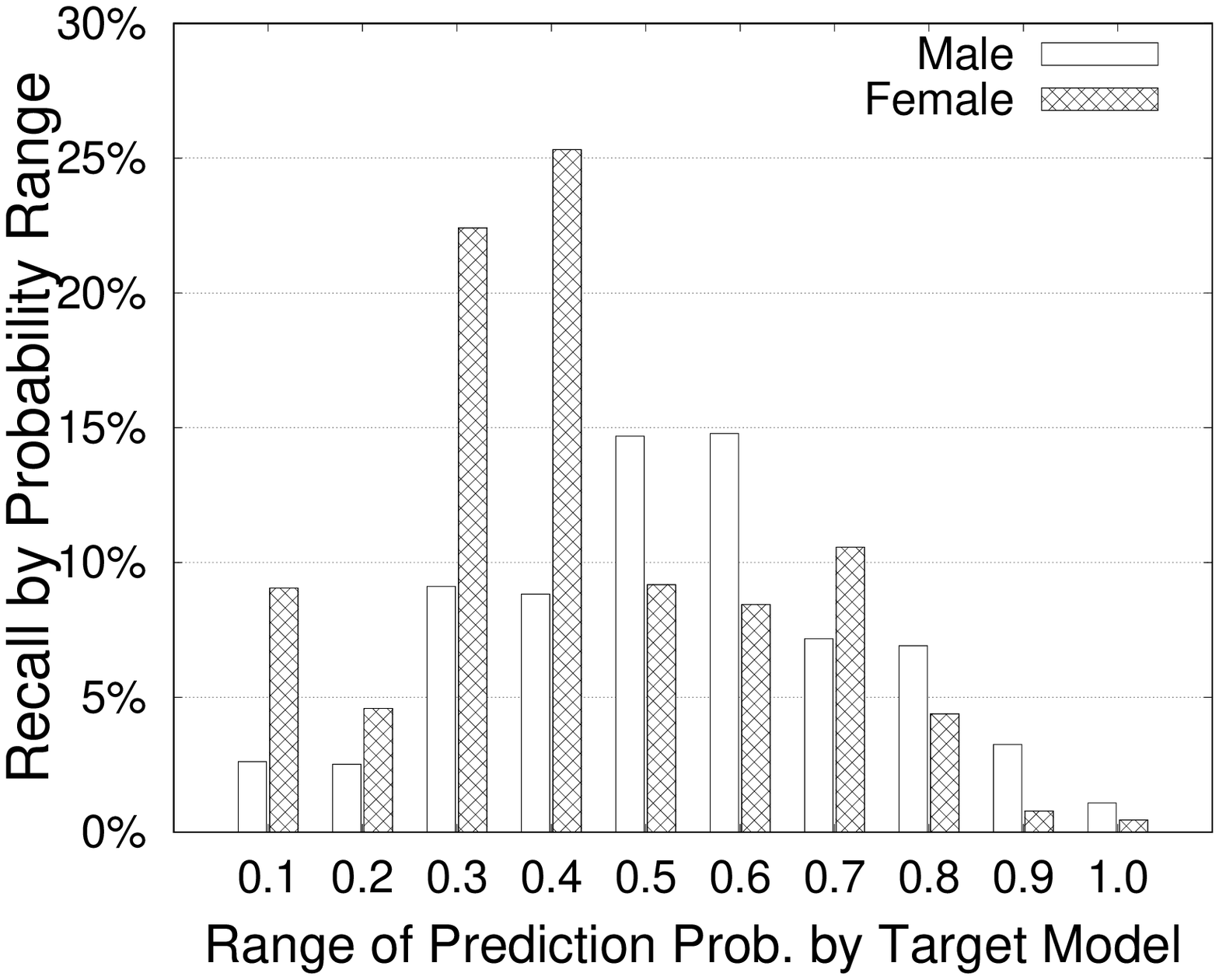}
          &
          \includegraphics[width=0.2\textwidth]{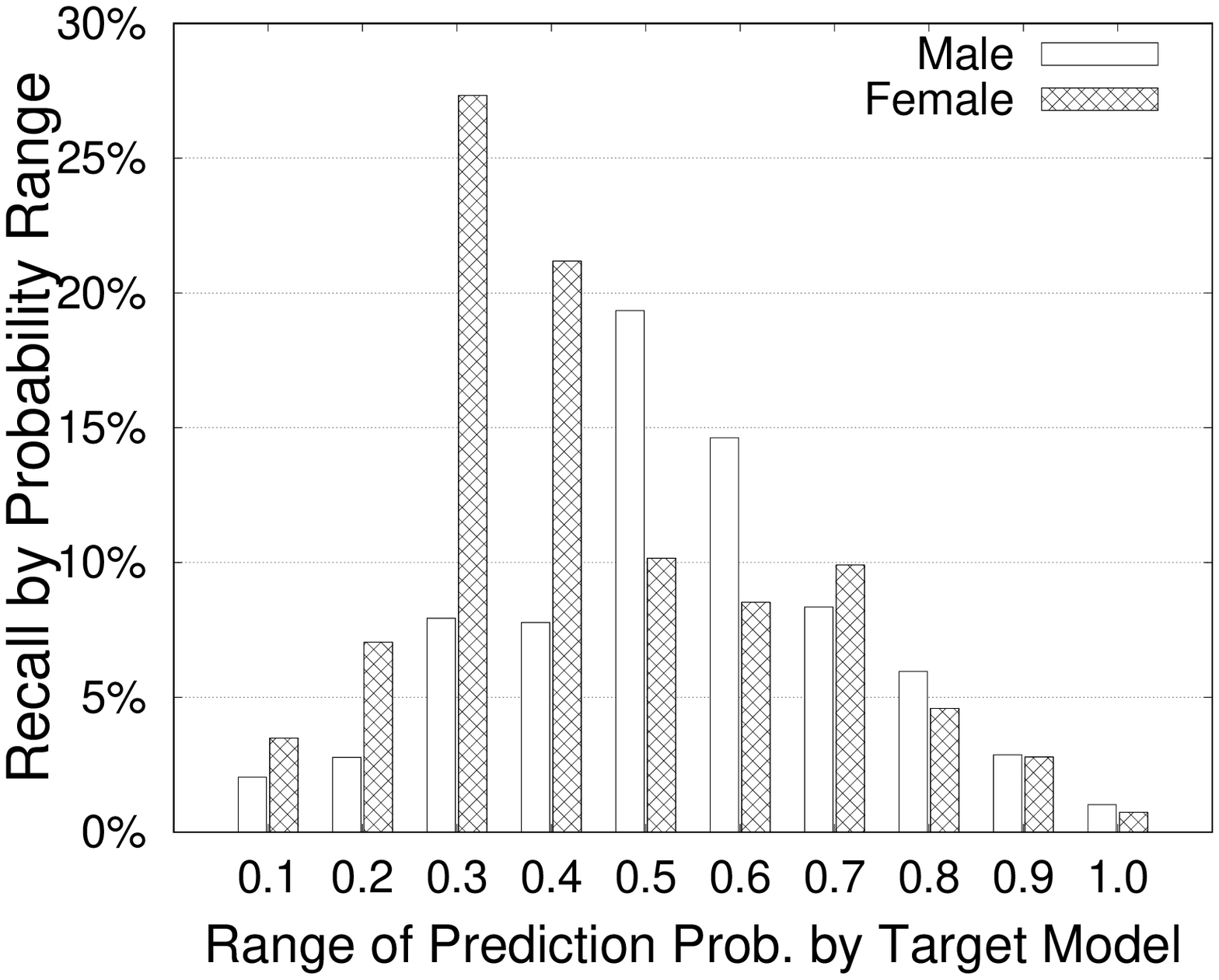}
          &
          \includegraphics[width=0.2\textwidth]{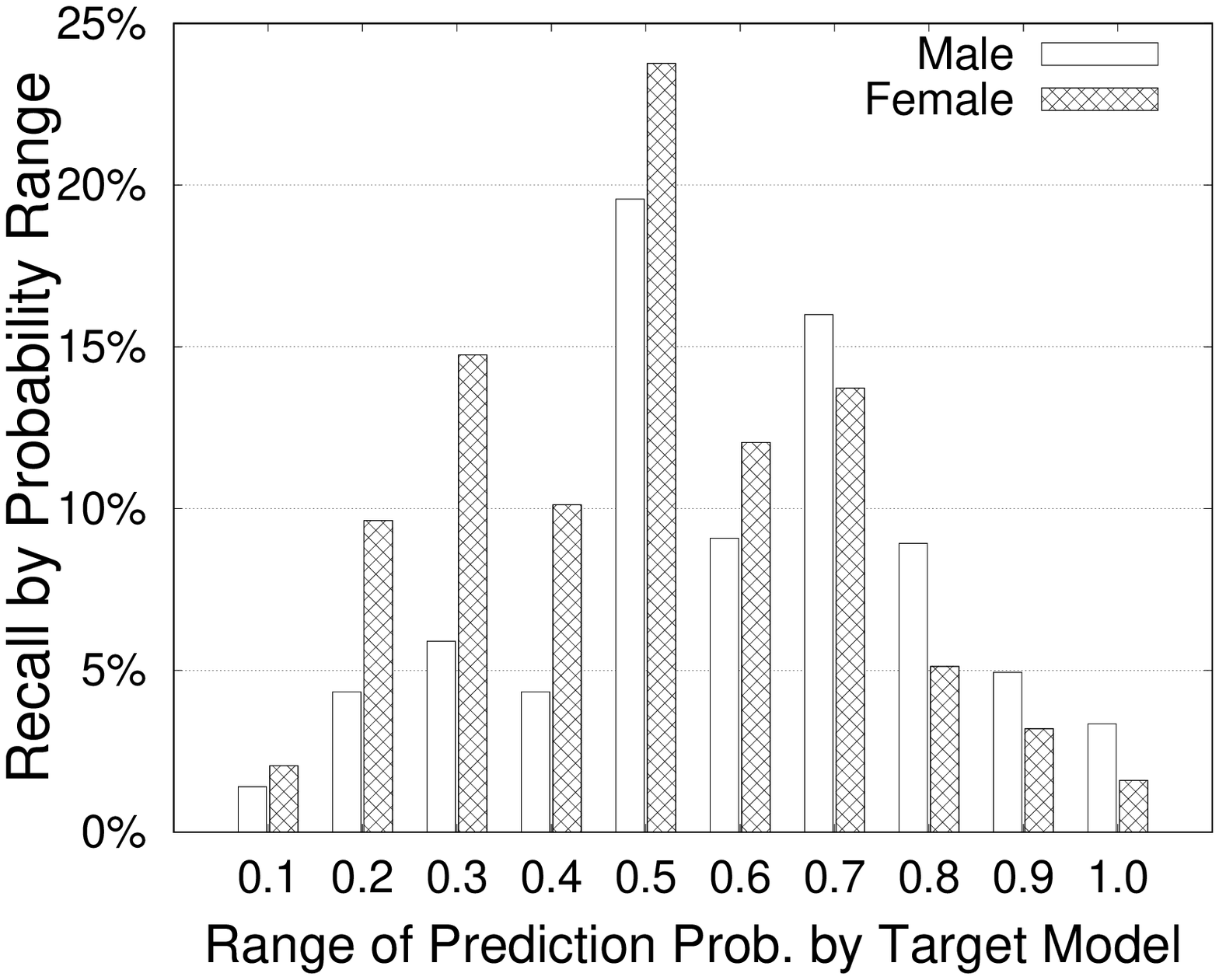}
          \\
       {\scriptsize (f) COMPAS (gender, No DP)}
        &
        {\scriptsize (g) COMPAS (gender, $\epsilon=1$)}
         &
        {\scriptsize (h) COMPAS (gender, $\epsilon=0.5$)}
         &
        {\scriptsize (i) COMPAS (gender, $\epsilon=0.1$)}
         &
        {\scriptsize (j) COMPAS (gender, $\epsilon=0.05$)}
        \\
         \includegraphics[width=0.2\textwidth]{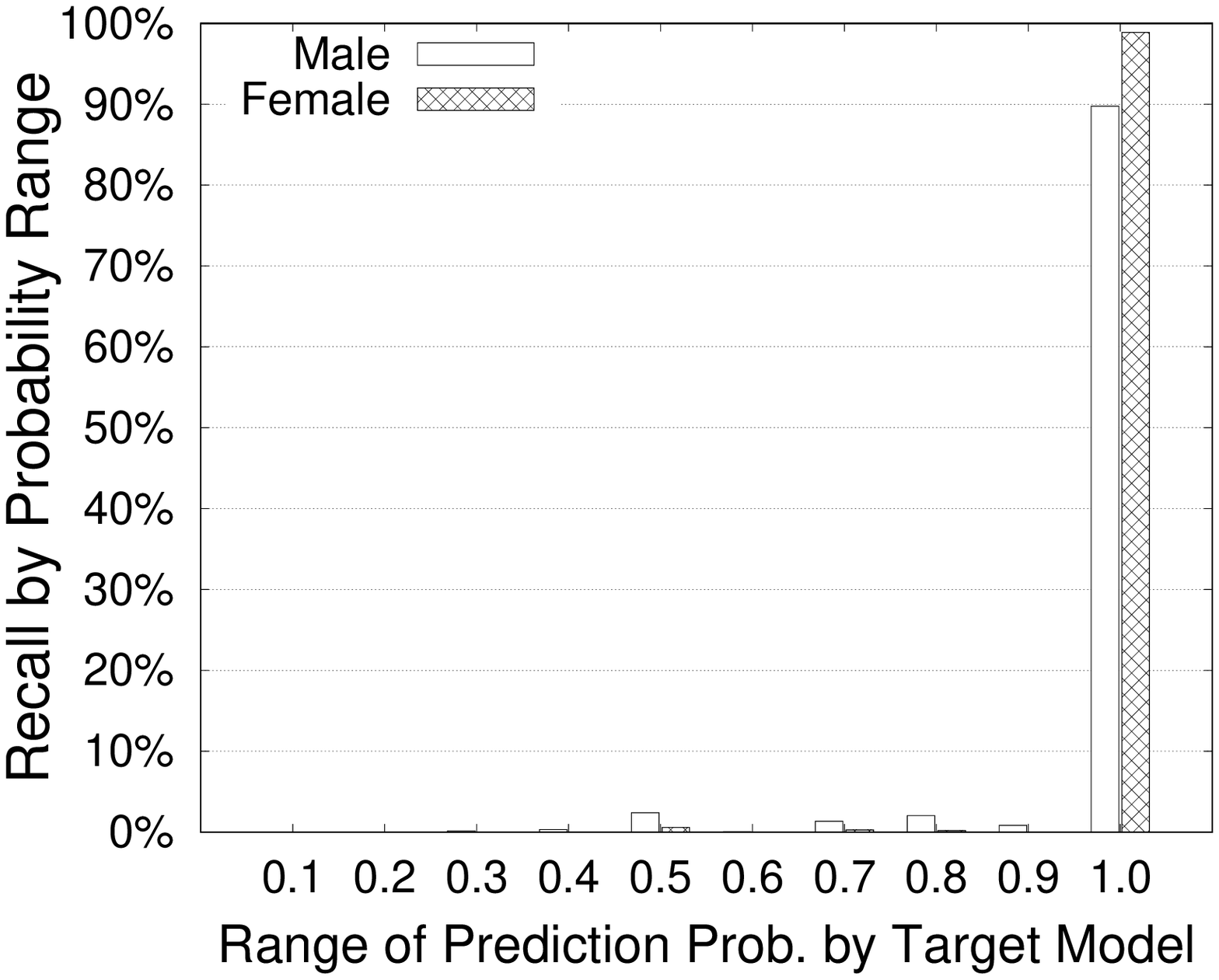}
          &
          \includegraphics[width=0.2\textwidth]{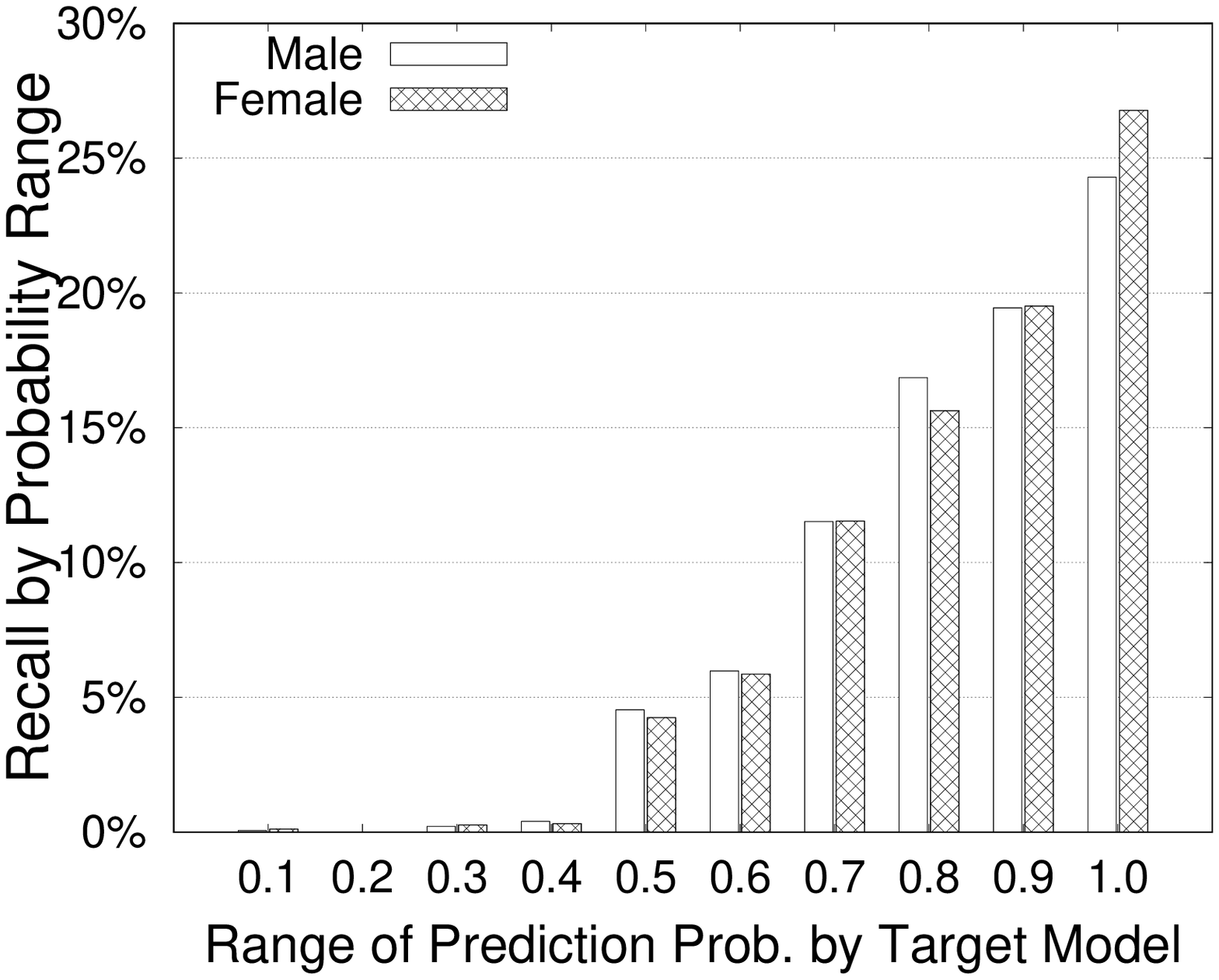}
           &
          \includegraphics[width=0.2\textwidth]{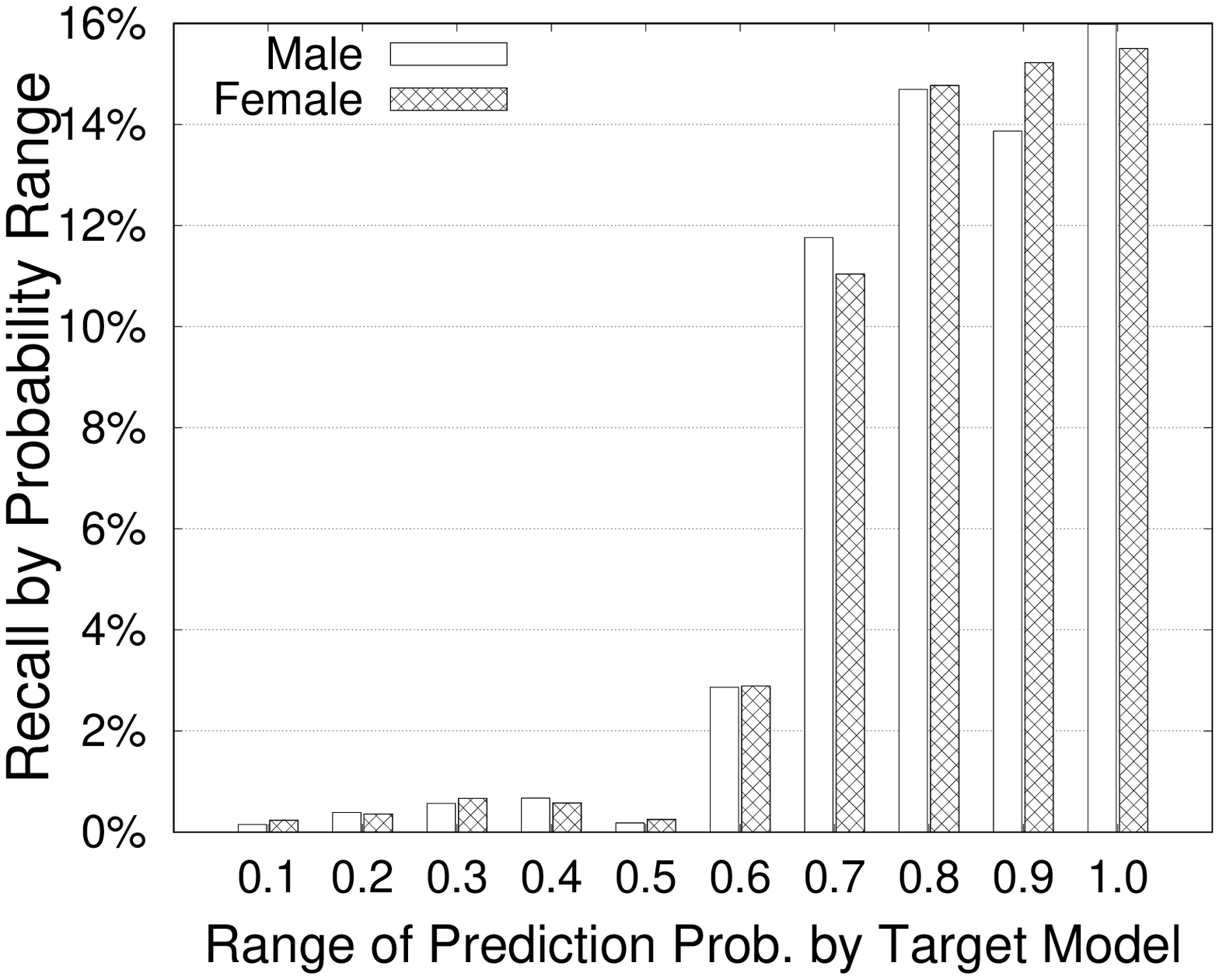}
          &
          \includegraphics[width=0.2\textwidth]{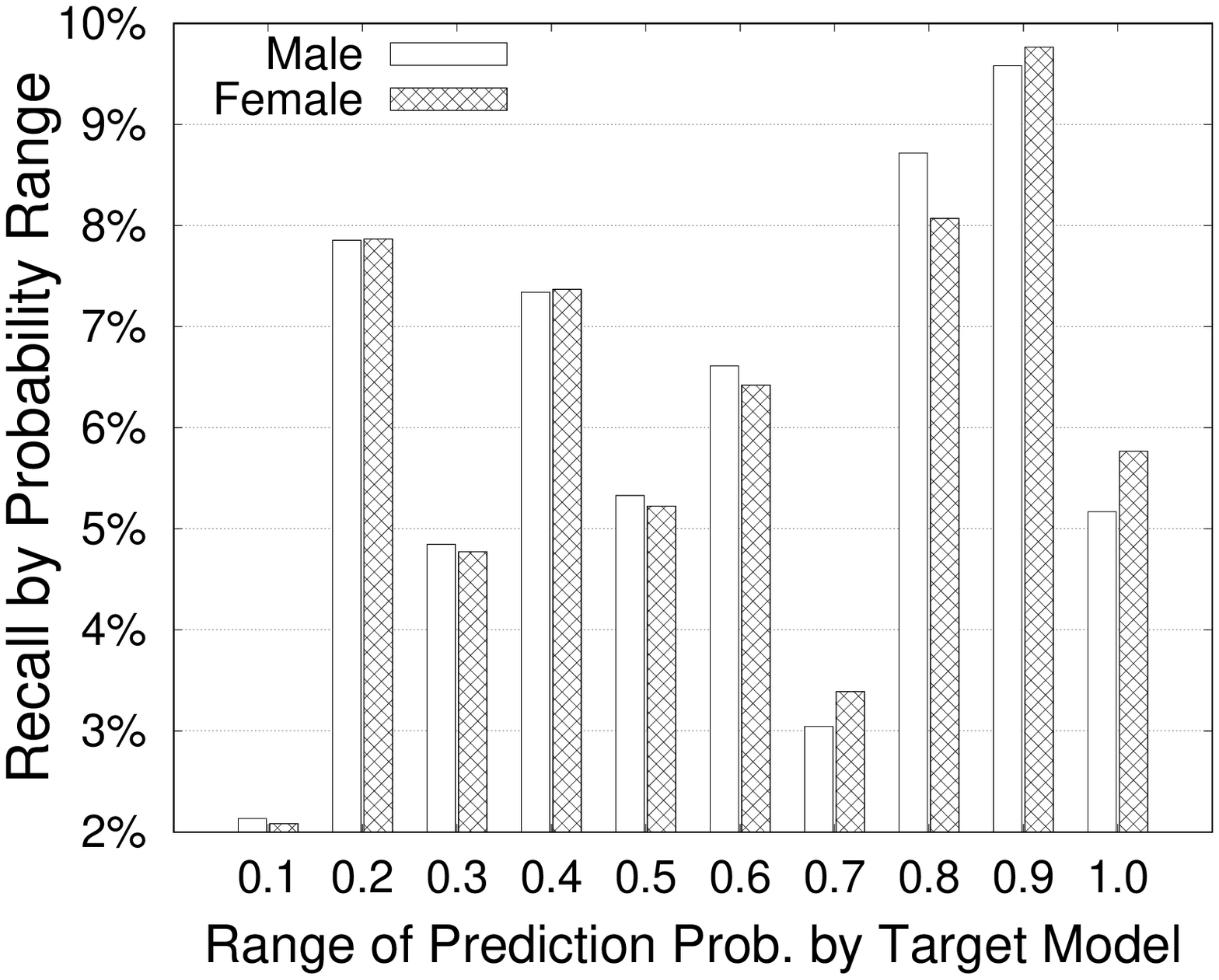}
          &
          \includegraphics[width=0.2\textwidth]{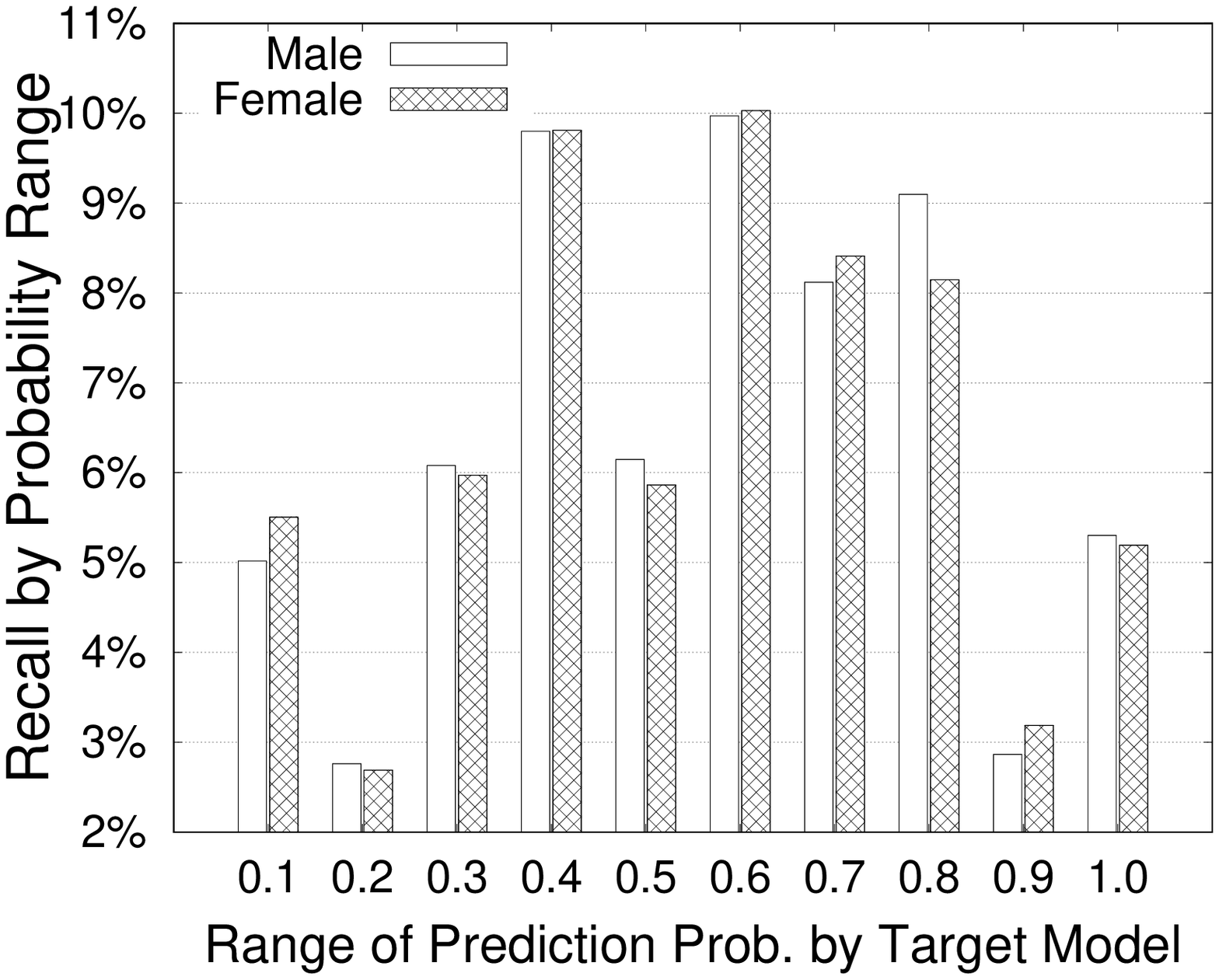}
          \\
            {\scriptsize (k) ADULT (gender, No DP)}
        &
        {\scriptsize (l) ADULT (gender, $\epsilon=1$)}
         &
        {\scriptsize (m) ADULT (gender, $\epsilon=0.5$)}
         &
        {\scriptsize (n) ADULT (gender, $\epsilon=0.1$)}
         &
        {\scriptsize (o) ADULT (gender, $\epsilon=0.05$)}
        \\
         \\
        \includegraphics[width=0.2\textwidth]{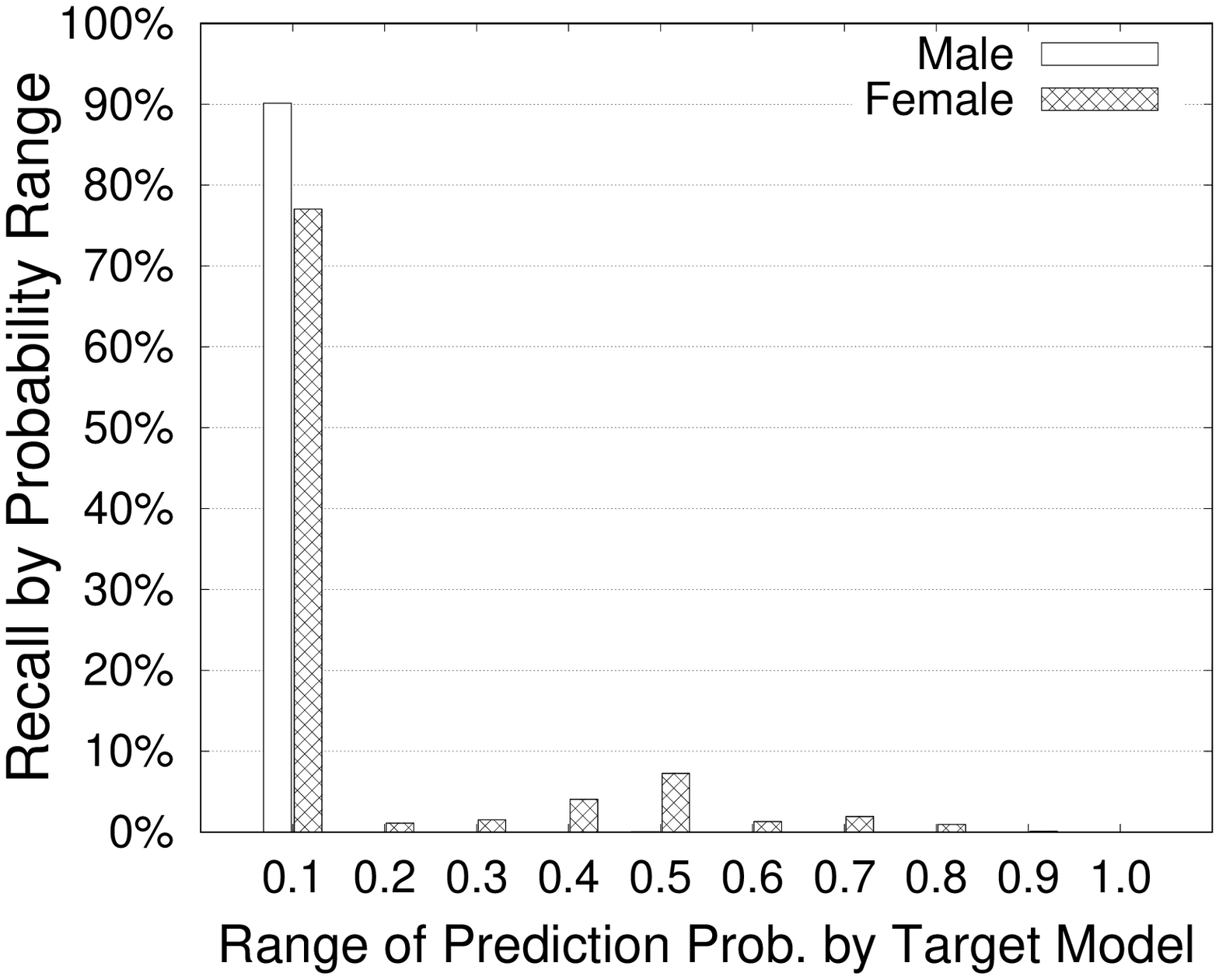}
          &
         \includegraphics[width=0.2\textwidth]{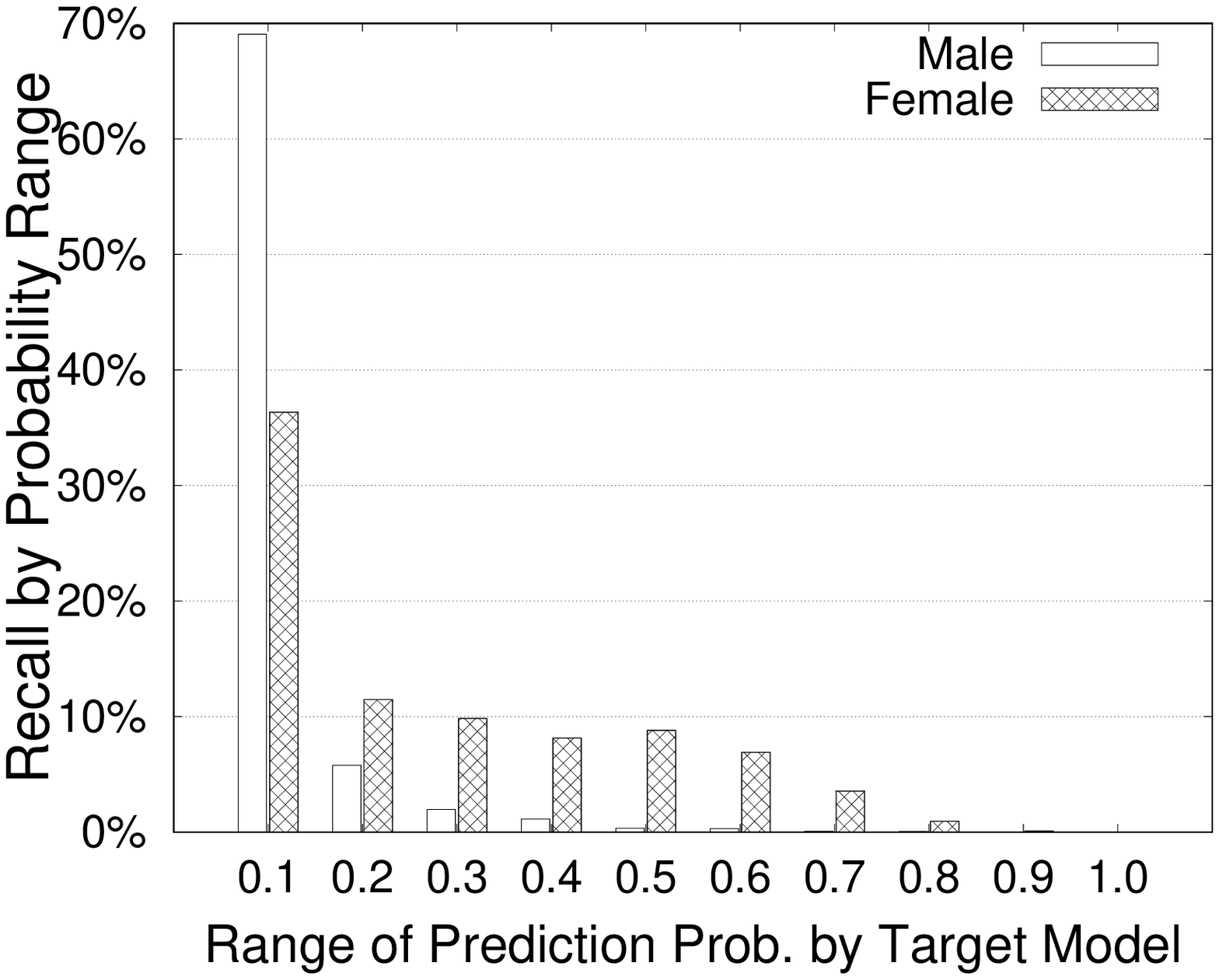}
           &
        \includegraphics[width=0.2\textwidth]{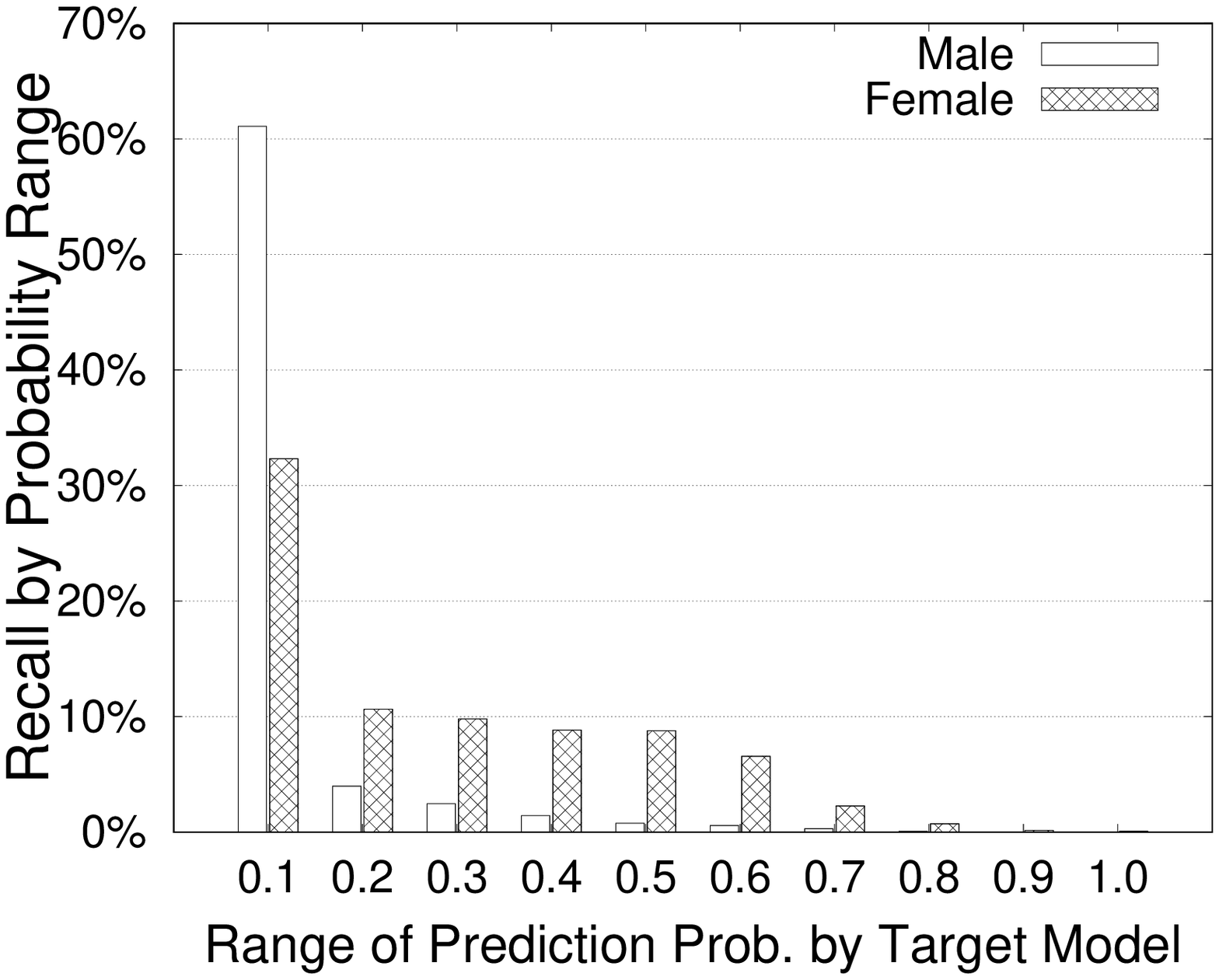}
          &
         \includegraphics[width=0.2\textwidth]{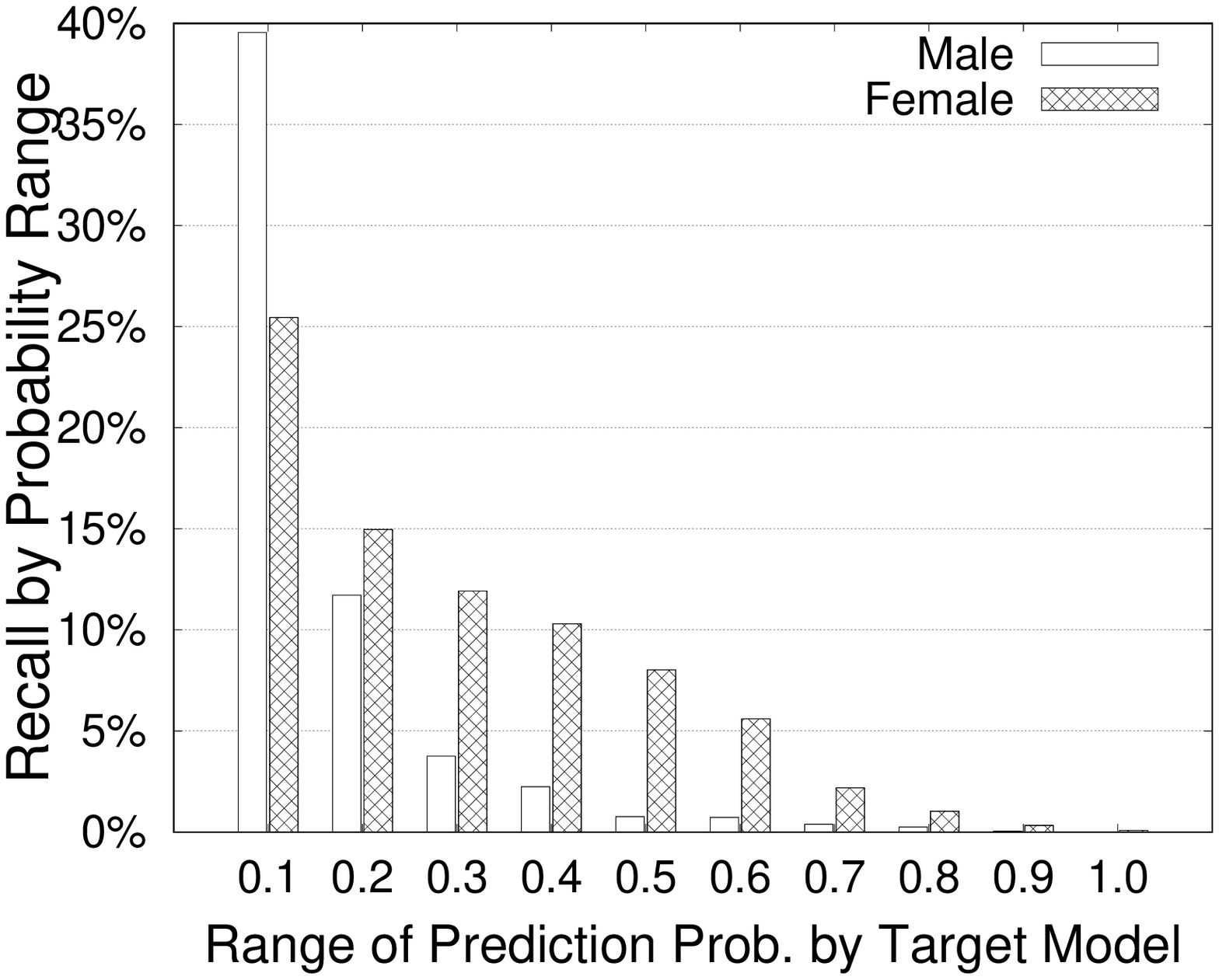}
          &
         \includegraphics[width=0.2\textwidth]{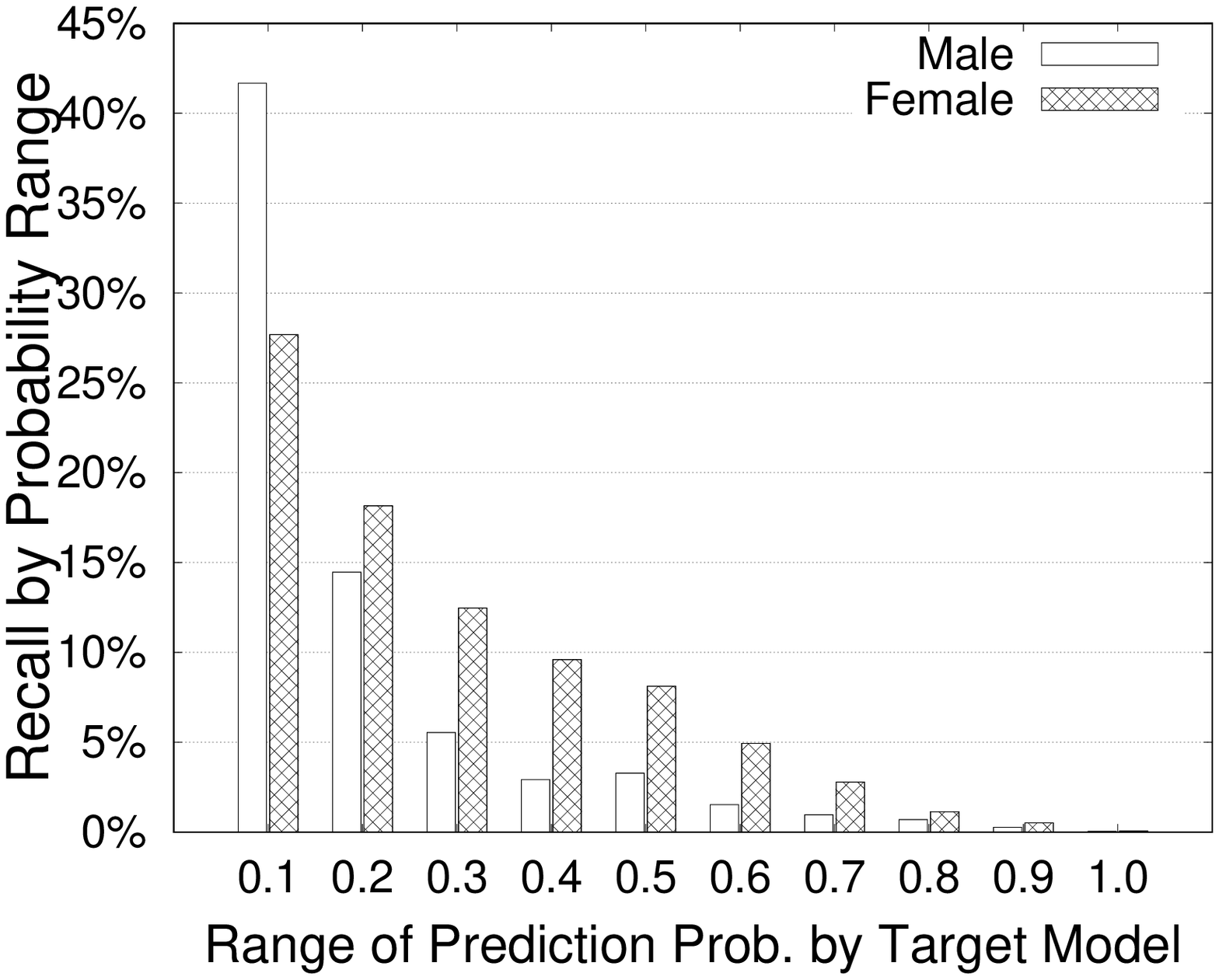}
          \\
            {\scriptsize (p) Hospital (gender, No DP)}
        &
        {\scriptsize (q) Hospital (gender, $\epsilon=1$)}
         &
        {\scriptsize (r) Hospital (gender, $\epsilon=0.5$)}
         &
        {\scriptsize (s) Hospital (gender, $\epsilon=0.1$)}
         &
        {\scriptsize (t) Hospital (gender, $\epsilon=0.05$)}
        \end{tabular}
       \caption{\small \label{fig:unsafe-range2} Recall of different groups by prediction probability range}
\end{figure*}

Next, we explain more details of our observations. We will provide  detailed analysis of the underlying reasoning behind these observations in Section \ref{sc:analysisvd}. 

 {\bf VD measurement before DP deployment.} We make one  interesting observation from Figure~\ref{fig:vul-disp} (for the case of ``No DP'').  
 In all datasets, the smaller group is not always more vulnerable against MIA than the larger group. For example, as shown in Figure \ref{fig:vul-disp} (b), the male group is always more vulnerable than the female group in Broward \& Adult datasets, while the female group is more vulnerable than the male group in COMPAS and Hospital datasets. This shows that VD is not directly related to the group distribution of the protected attributes alone. 

 {\bf VD measurement after DP.} As shown in Figure~\ref{fig:vul-disp},  our results support the conclusion that \emph{DP cannot eliminate VD}. Furthermore, the strength of DP (controlled by using various privacy budgets) has some clear impacts on VD. However, the change of VD does not change linearly with the privacy budget.  

{\bf Change of VD by DP.} The results (Figure \ref{fig:vul-disp-against-dp}) do not show a consistent pattern of VD reduction when the privacy budget increases. Furthermore, DP does not always reduce the amounts of VD. 
On Hospital dataset DP even increases VD (as shown in Figure~\ref{fig:vul-disp} (d)) from close to 0 to as large as 0.2.

\subsection{Analysis of Vulnerability Disparity} 
\label{sc:analysisvd}
In this section, we explain the reasons behind our main observations in Section \ref{sc:vdmeasurement}. 

In a MIA attack, both target and attack models together create a mapping from the input domain to MIA predictions. Remind that the input of MIA, which is also the output of the target model, consists of a vector of prediction probabilities, each probability measuring if an input data point  belongs to a specific class $i$. In essence, MIA maps each prediction probability to a binary label.   
Certain ranges of prediction probabilities are projected to the positive label by MIA (i.e., predicted as members). 
Intuitively, VD is directly related to how the training samples from different groups are mapped to the binary label.  
Recall that VD indeed measures the difference of recall of different groups. 
Following the above reasoning, we performed a set of experiments to measure the recall of different groups 
before and after DP is enforced on the target model.  
To better illustrate our results, we evenly split the probability domain [0, 1] into 10 ranges, each of size 0.1. 
For each range $r$, we count the number of records (denoted as $c_r^{A=a}$) in group $A=a$ that are labeled with the positive label by MIA and their prediction probability falls into $r$. Then we calculate the {\em recall of a group $A=a$ in the range $r$}, denoted as $Rec_r$, as 
\begin{equation}
Rec_r^{A=a}= \frac{c_r^{A=a}}{c^{A=a}},
\end{equation}
where $c^{A=a}$is the number of records in group $A=a$ that are labeled with the positive label by MIA. Apparently, the difference $(Rec_r^{A=a} - Rec_r^{A=\bar{a}})$ between the recall of the two groups in the range $r$ equals the amount of VD in $r$. Also it must be true that: 
\[VD = \sum_{i=1}^{10} (Rec_{r_i}^{A=a} - Rec_{r_i}^{A=\bar{a}}). \]
In other words, the sum of the recall difference in each range should equal to VD of the whole dataset.  We illustrate the results of recall by probability range in Figure \ref{fig:unsafe-range2}. 

\noindent{\bf Explanations.} Next, We will answer the following three questions raised from our observations in Section \ref{sc:vdmeasurement} by analyzing how the recall of different groups vary, and how the recall per group shifts with regards to different degrees of privacy protection (controlled by privacy budget $\epsilon)$. We vary $\epsilon$ from 0.1 to 5 (i.e., from strong to weak privacy protection). 

{\bf Why does VD exist?}
As shown in Figure \ref{fig:unsafe-range2} (a), (f), (k) and (p), before the deployment of DP, various groups have different recall in each probability range. For instance, consider Broward dataset (Figure \ref{fig:unsafe-range2} (a)), the true positive records  mostly cover the prediction probabilities in the range of [0, 0.1), in which the recall of male and female groups differ around 10\%. This leads to the VD (around 0.1) of the two groups (Figure \ref{fig:vul-disp-against-dp} (b)). The same observation also holds on the other three datasets. We have to note that on ADULT dataset, the positive records mainly cover two ranges of prediction  probabilities, [0, 0.1) and [0.9, 1]. These two ranges show the opposite recall results: the recall of the male group is higher than the female group in the range [0, 0.1) while opposite in the range [0.9, 1]. Also the recall difference in these two ranges is of similar amounts. Therefore, the VD on ADULT dataset is relatively small (as shown in Figure \ref{fig:vul-disp} (a)). 

{\bf What determines the amount of VD?}
Since the target model determines such recall difference, we next investigate if the  gender feature is the main factor. We identify the important features of Broward dataset, which are those features that directly effect the output probability vector of the target model. It turns out that the top-3 important features of Broward dataset is {\em Decile}, {\em Priors\#}, and {\em Age}. We studied two data distributions: (1) data distribution $D_1$ on the combination of the four features (i.e., the top-3 important features plus {\em gender}); and (2) data distribution $D_2$ on the top-3 important features. 
We compared $D_1$ and $D_2$, and found out that the difference between $D_1$ and $D_2$ is closely (not completely) consistent with the recall distribution shown in Figure \ref{fig:unsafe-range2} (a). The same observation also holds on other datasets. For example, on ADULT dataset, whose top-3 important features are {\em Education}, {\em Marital Status}, and {\em Cap\_Gain}, shows almost identical distributions on the combination of top-3 important features alone and the top-3 features plus {\em gender}. This leads to the small VD on Adult dataset (Figure \ref{fig:vul-disp} (a)). In summary, the amount of VD is not determined by the data distribution on the protected attribute alone. Instead, it is determined by the data distribution on the combination of the protected attribute with other attributes (sometimes the top-3 important features are sufficient).

{\bf How does DP impact VD?} As shown in Figure \ref{fig:unsafe-range2}, on all the four datasets, introducing noise to the target model makes fundamental impacts on its output. Specifically, smaller privacy budget (i.e., stronger privacy protection) introduces more randomness into the target model, and thus more unique prediction probabilities in the output of the target model. These probabilities scatter more widely in the domain. 
Meanwhile, from the perspective of individual records, due to the  noise added by DP, the prediction probability of two identical records can be different. Therefore, smaller privacy budget leads to higher likelihood that two identical records are output as  different prediction probabilities by the target model. However, our analysis shows that there is no linear correlation between the privacy budget and the size of unsafe areas (i.e., the number of prediction probabilities that are mapped to the positive label by MIA), due to the randomness of perturbation on the target model  added by DP. 
In other words, the mapping from the input domain to MIA predictions becomes probabilistic. 
Therefore, the amount of VD is unpredictable and change inconsistently with the increment of the privacy budget. This has been demonstrated by the results in Figures \ref{fig:vul-disp} and  \ref{fig:vul-disp-against-dp}. This is expected since DP, as a privacy notion, does not deal with the discriminated treatment by MIA (indeed any machine learning model) by nature. Furthermore, since the DP enforcement on the target model does not consider the protected attributes as well as the data distribution on these attributes, it is anticipated that DP does not lead to consistent change of VD on these protected attributes at all. 
\section{Mitigation Algorithms for Vulnerability Disparity}
\label{sc:mitigation}
As is illustrated in Section~\ref{sc:analysisvd}, the vulnerability disparity is caused by inconsistent distribution of training samples from different groups. To fundamentally solve this problem, generally speaking, there are two approaches. First, we can ``optimize'' the target model such that, given the training samples from different groups as inputs, it produces equally-distributed outputs. Second, we can elaborately pick similarly-distributed data from different groups to train the target model. The first approach, however, has several major limitations. First, there lacks theoretical support of designing such a model. It remains an open question whether we can achieve our goal while preserving the properties of the original model. Second, it needs separate solutions for different types of target models. This brings high complexities to generalize the approach. Finally, it will be incompatible with other defenses on top of the target model (\eg, the DP we used in this study) since those defenses can reversely vary the output distribution. Therefore, in this research, we choose to explore the second approach.

\subsection{Design of Vulnerability Correction}
\label{subsec:fairpick}
Past research has proposed several data pre-processing approaches for addressing unfairness issues~\cite{kamiran2012data}. In particular, reweighing~\cite{calders2009building} and its variants~\cite{kamiran2012data} are widely used to facilitate classification without discrimination. Technically speaking, reweighing picks training samples from every group such that they share identical distribution as the entire data-set in the dimension of label. That is to say, reweighing makes all groups have consistent distribution when we consider the label as the index. By intuition, we can extend reweighing to achieve our goal by considering the feature combination as the label. This approach, however, can have two problems because of reweighing's strict requirement of distribution. First, reweighing frequently needs to over-pick (or duplicate) samples from a group to meet the needs of distribution. However, from a practical perspective, the over-picked samples do not truly affect the distribution of real samples. Therefore, reweighing may not provide realistic protection. Second, the feature combinations are often sparsely distributed, in particular when the features are many while the data-set is small. As a consequence, plenty of feature combinations only appear in a specific group and to ensure consistent distributions, reweighing has to delete samples with those feature combinations. This frequently leads to over-deletion and makes the resulted data-set unusable. 

\begin{algorithm}[!t]
\caption{Algorithm for \tool{}}
\label{alg:delete-reweigh}
\begin{algorithmic}[1]
\STATE{ {\bf Input} $D_{train}$: Private training data-set; $\mathcal{S}$: Protected attribute domain; $T$ reduction rate;}
\STATE{ {\bf Let} $D'_{train}$ = $\emptyset$ to be the training data-set after \tool{}}
\STATE{ Split $D_{train}$ by the class label and get a set of data-sets $C_{train}$}
\STATE{ {\bf for} each $d \in C_{train}$ {\bf do}}
\STATE{\ \ \ \ $d_{kmeans} = Kmeans(d)$}
\STATE{\ \ \ \ In each cluster, calculate its center to construct a new feature combination.}
\STATE{\ \ \ \ {\bf Let} $f$ to be the all the feature combination of $d_{kmeans}$}
\STATE{\ \ \ \ {\bf Let} $func.=\emptyset$}
\STATE{\ \ \ \ {\bf for} i = 1\ :\ |S| {\bf do}}
\STATE{\ \ \ \ \ \ \ \ \ \ \ \ create Equation \ref{eq:reduce-com2} of $del(G_i, C_j) (1\le j \le t)$ for $G_i$}
\STATE{\ \ \ \ \ \ \ \ \ \ \ \ Append the equations to $func$}
\STATE{\ \ \ \ {\bf end for} }
\STATE{\ \ \ \ ans = quadratic.programming($func$)}
\STATE{\ \ \ \ $d_{reweigh}$ = pick\_samples(ans, $d_{kmeans}$)}
\STATE{\ \ \ \ $d'$ = Map records in $d_{reweigh}$ to original records}
\STATE{\ \ \ \ Append $d'$ to $D'_{train}$}
\STATE{{\bf end for}}
\STATE{{\bf return} $D'_{train}$}
\end{algorithmic}
\end{algorithm}

\noindent{\bf Our approach.} To address the above problems, we propose \tool{}. Instead of seeking fully consistent distribution among different groups, \tool{} reduces the distribution difference by a pre-defined threshold. Formally, we consider a data-set $D$ and a protected attribute $F_p$ which defines $n$ groups $\{G_1, G_2, ..., G_n\}$. Beyond $F_p$, the data has a group of other attributes $\{F_1, F_2, ..., F_m\}$. From all records in $D$, $\{F_1, F_2, ..., F_m\}$ has $t$ unique combinations $\{C_1, C_2, ..., C_t\}$. With $sum(G_i)$ indicating the total number of training samples in group $G_i$ and $sum(G_i, C_j)$ referring to the number of training samples with combination $C_j$ in group $G_i$, the distribution difference on $C_j$ between $G_i$ and other groups can be defined as:

\begin{equation}
\begin{aligned}
\label{eq:reduce-com1}
dvar(G_i,C_j) = \frac{sum(G_{i}, C_{j})}{sum(G_{i})} - \frac{\sum^{k \neq i} sum(G_{k}, C_{j})}{\sum^{k \neq i} sum(G_{k})}
\end{aligned}
\vspace{2pt}
\end{equation}

\noindent After applying \tool{}, we require that: 

\begin{equation}
\begin{aligned}
\label{eq:reduce-com2}
\forall i \in [1,n],\forall j \in [1,t],\quad dvar(G_i, C_j, post)= dvar(G_i, C_j, pre) * T
\end{aligned}
\vspace{2pt}
\end{equation}

\noindent where  $dvar(G_i, C_j, pre)$ and  $dvar(G_i, C_j, post)$ respectively represent $dvar(G_i, C_j)$ before and after \tool{}, and $T$ ($0 \le T \le 1$) is a user-specified threshold. To better understand the effects of Equation~\ref{eq:reduce-com2} on VD, let's assume that the unsafe areas contain $\{C_{u1}, C_{u2}, ..., C_{uk}\}$ (which will be determined as training samples by MIA). For simplicity, we also assume the unsafe areas do not shift after \tool{}. Under such contexts, VD pertaining to group $G_i$, before and after we apply \tool{}, respectively equals to $\sum_{n=u1}^{uk}dvar(G_i,C_j,pre)$ and  $\sum_{n=u1}^{uk}dvar(G_i,C_j,post)$. According to Equation~\ref{eq:reduce-com2}, it is not hard to observe that VD will be reduced by $(1-T)$.  

To solve Equation~\ref{eq:reduce-com2}, we consider deleting $del(G_i, C_j)$ ($0 \le j \le t$,$0 \le i \le n$) samples that have feature combinations $C_j$  from group $G_i$. This creates $t * n$ variables. Including $del(G_i, C_j)$ to Equation~\ref{eq:reduce-com1}, we will have $dvar(G_i,C_j,post)$ equal to: 

\begin{equation}
\begin{aligned}
\label{eq:reduce-com3}
\frac{sum(G_{i}, C_{j}) - del(G_i, C_j)}{sum(G_{i}) - \sum_{n=1}^{t}del(G_i, C_n)} - \frac{\sum^{k \neq i} (sum(G_{k}, C_{j})-del(G_k, C_j))}{\sum^{k \neq i} (sum(G_{k}) -  \sum_{n=1}^{t} del(G_k, C_n))}
\end{aligned}
\vspace{2pt}
\end{equation}

\noindent By correlating Equation~\ref{eq:reduce-com3} with Equation~\ref{eq:reduce-com2}, we can establish $t * n$ equations with $del(G_i, C_j)$ ($0 \le j \le t$,$0 \le i \le n$) as variables. While it is hard to derive general solutions, these equations are typically solvable with quadratic programming~\cite{frank1956algorithm}.

\begin{figure*}[h]
\vspace{-0.1in}
\centering
\begin{tabular}{@{}c@{}}
        \begin{tabular}{@{}c@{}c@{}c@{}c@{}c@{}}
          \includegraphics[width=0.25\textwidth]{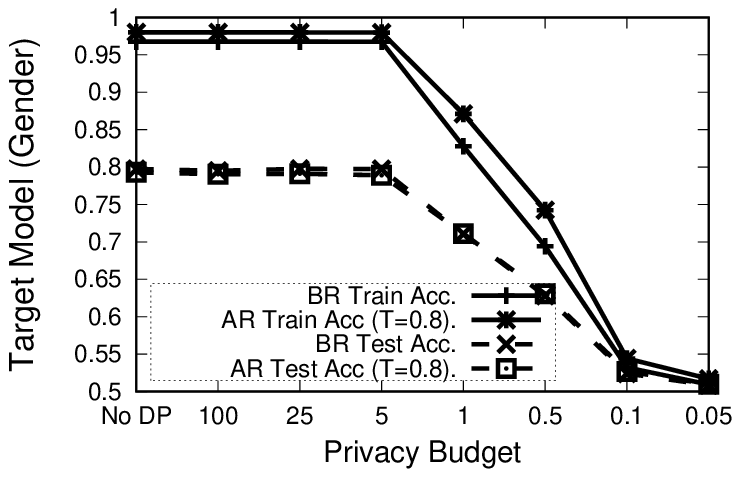}
           &
          \includegraphics[width=0.25\textwidth]{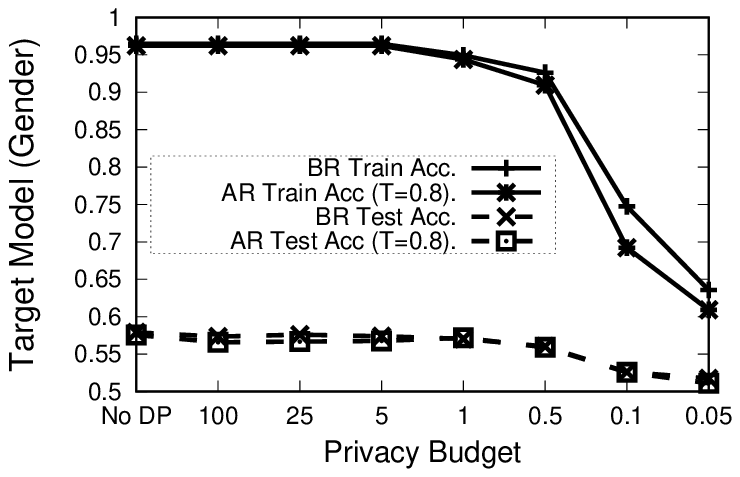}
          &
          \includegraphics[width=0.25\textwidth]{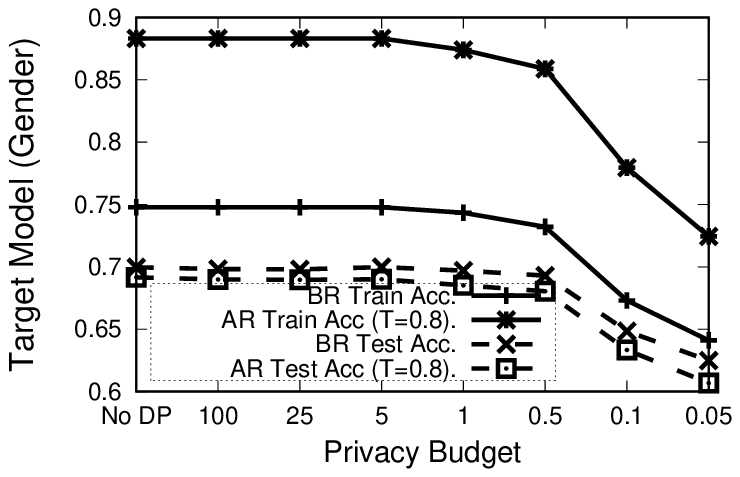}
          &
                    \includegraphics[width=0.25\textwidth]{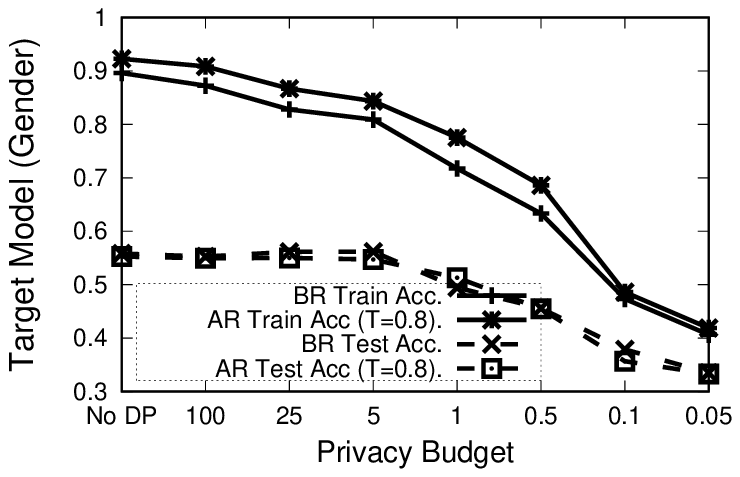}
          
        \end{tabular}
        \\
        \begin{tabular}{@{}c@{}c@{}c@{}c@{}c@{}}
          \includegraphics[width=0.25\textwidth]{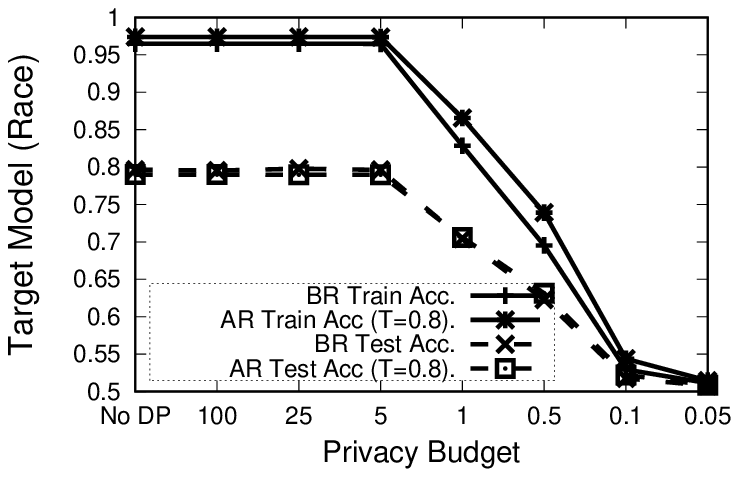}
           &
          \includegraphics[width=0.25\textwidth]{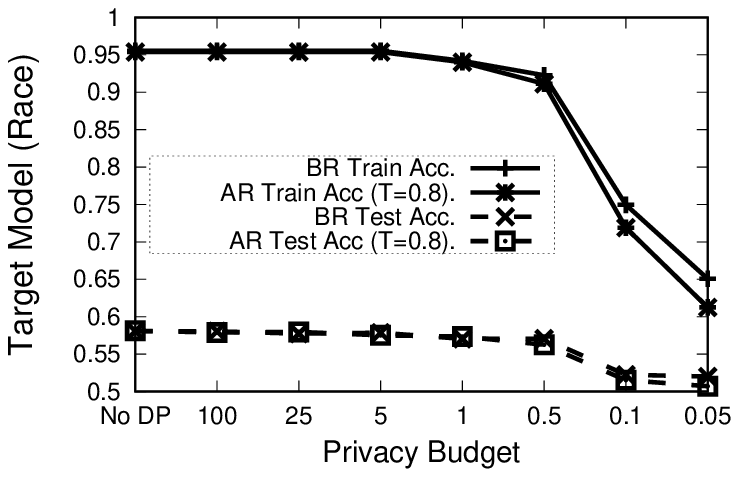}
          &
            \includegraphics[width=0.25\textwidth]{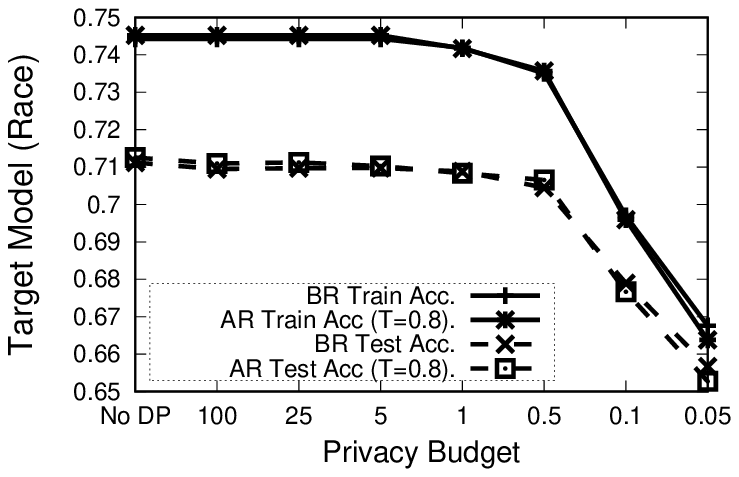}
            &
          \includegraphics[width=0.25\textwidth]{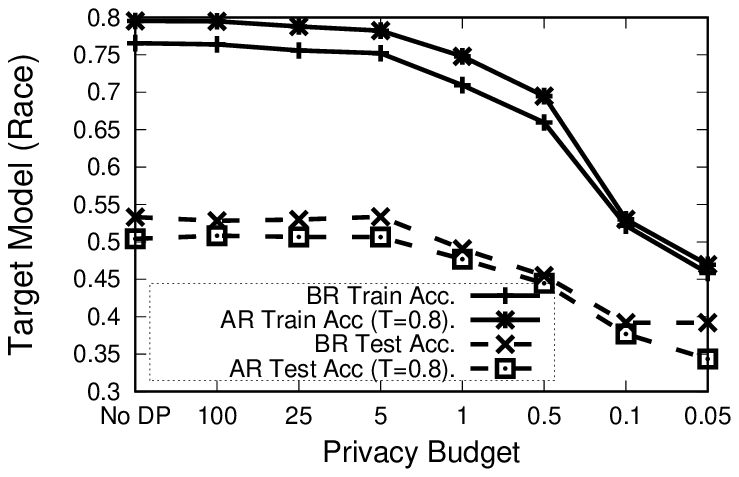}

        \end{tabular}
        \\
        \begin{tabular}{@{}c@{}c@{}c@{}c@{}c@{}}
          \includegraphics[width=0.25\textwidth]{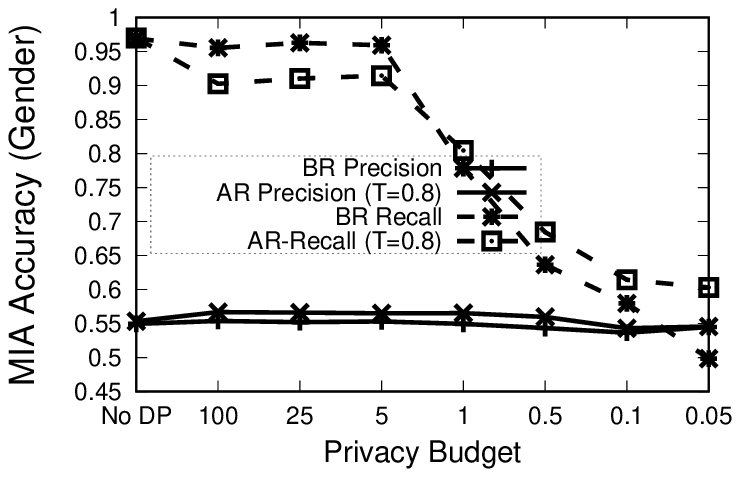}
           &
          \includegraphics[width=0.25\textwidth]{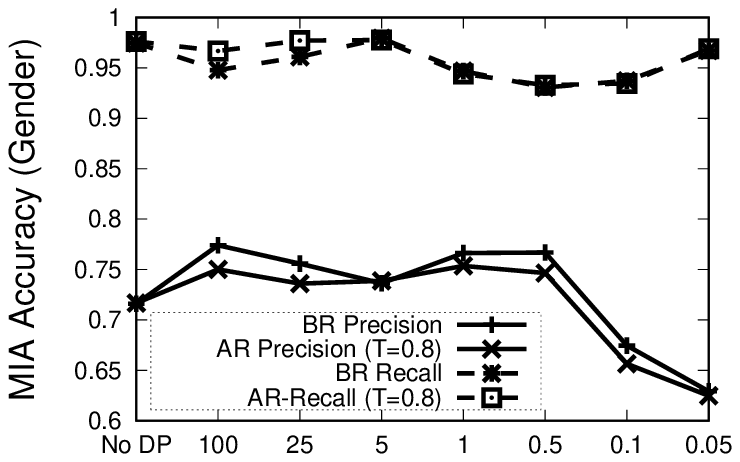}
          &
          \includegraphics[width=0.25\textwidth]{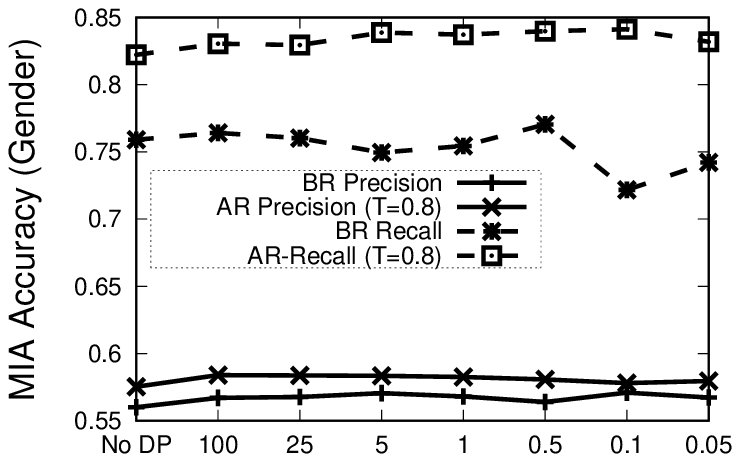}
          &
          \includegraphics[width=0.25\textwidth]{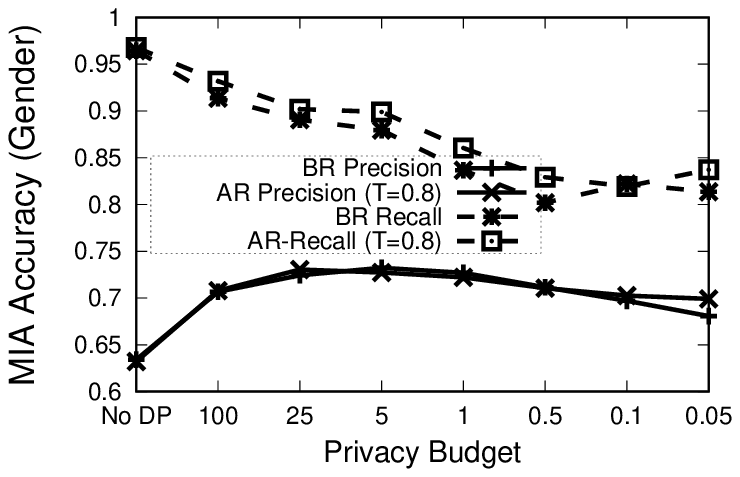}
          
        \end{tabular}
        \\
        \begin{tabular}{@{}c@{}c@{}c@{}c@{}c@{}}
          \includegraphics[width=0.25\textwidth]{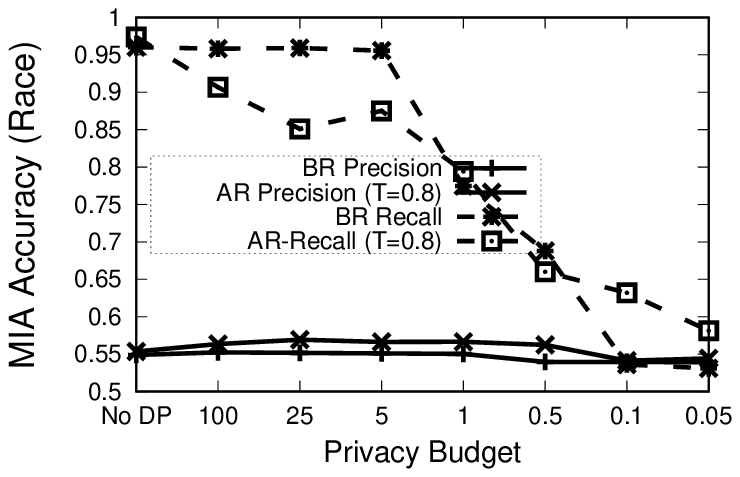}
           &
          \includegraphics[width=0.25\textwidth]{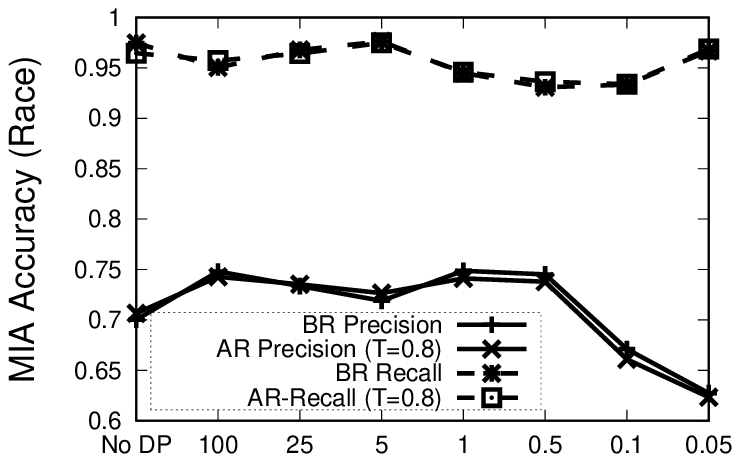}
          &
          \includegraphics[width=0.25\textwidth]{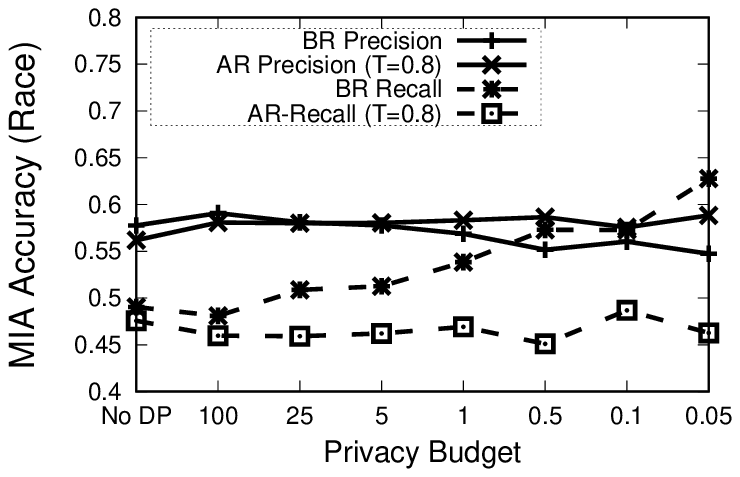}
          &
          \includegraphics[width=0.25\textwidth]{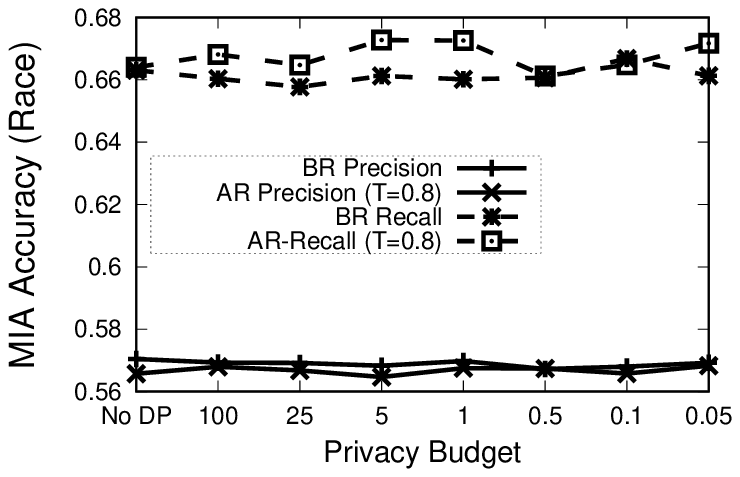}
          
        \\
       {\scriptsize (a) {\bf ADULT}}
        &
        {\scriptsize (b) {\bf BROWARD}}
         &
        {\scriptsize (c) {\bf COMPAS}}
         &
        {\scriptsize (d) {\bf Hospital}}
        \end{tabular}
\end{tabular}
        \vspace{-0.1in}
       \caption{\small \label{fig:target-model-acc-reweigh} Impacts of \tool{} in the target model and MIA (BR:Before reduction by \tool{}; AR: After reduction by \tool{})}
\end{figure*}

\begin{figure*}[!t]
\vspace{-0.1in}
\centering
        \begin{tabular}{@{}c@{}c@{}c@{}c@{}c@{}}
          \includegraphics[width=0.25\textwidth]{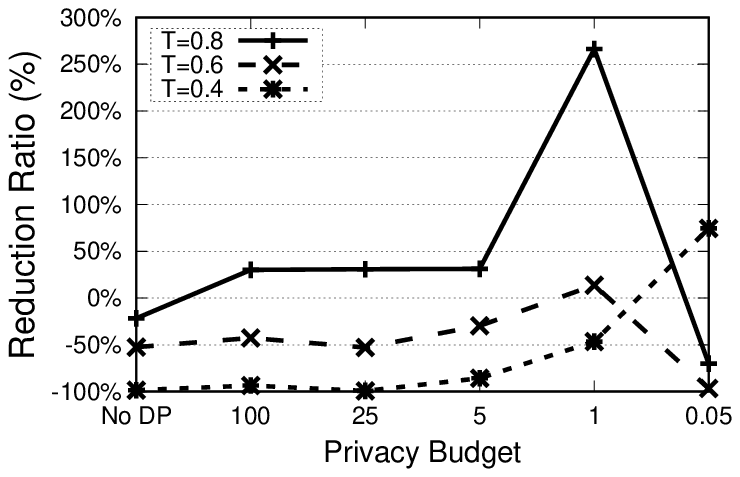}
           &
          \includegraphics[width=0.25\textwidth]{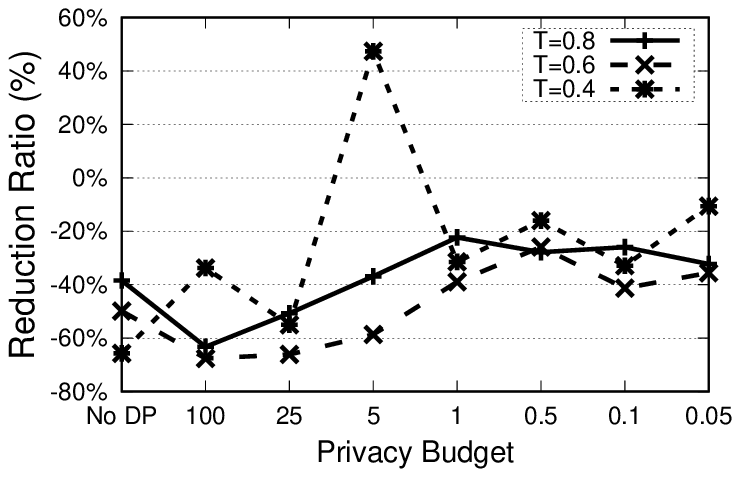}
          &
          \includegraphics[width=0.25\textwidth]{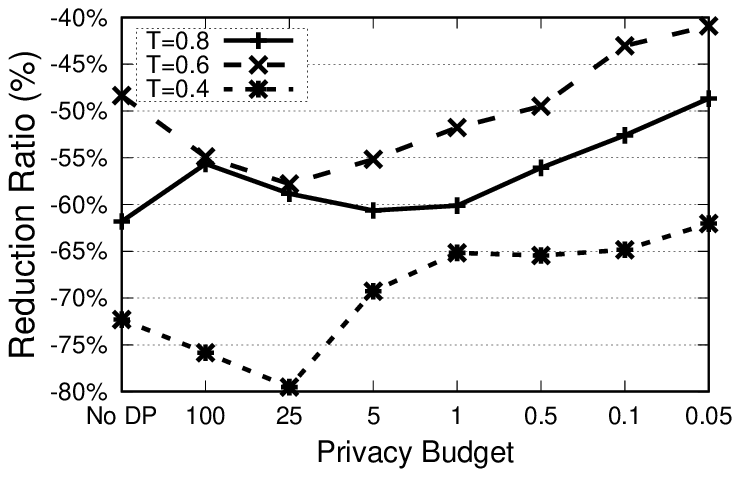}
           &
          \includegraphics[width=0.25\textwidth]{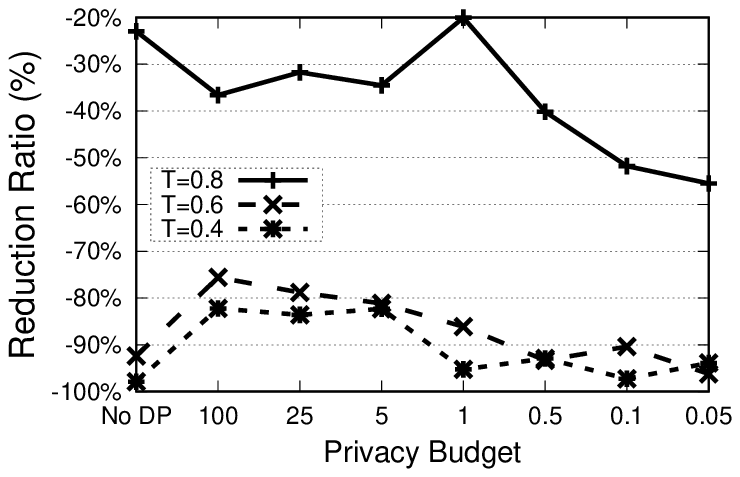}
        \\
        {\scriptsize (a) ADULT (Gender)}
        &
        {\scriptsize (b) BROWARD (Gender)}
         &
        {\scriptsize (c) COMPAS (Gender)}
         &
        {\scriptsize (d) Hospital (Gender)}
        \\
         \includegraphics[width=0.25\textwidth]{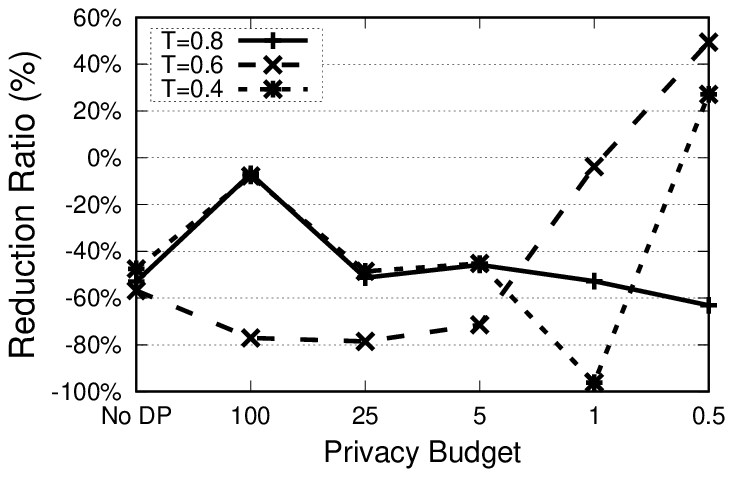}
           &
          \includegraphics[width=0.25\textwidth]{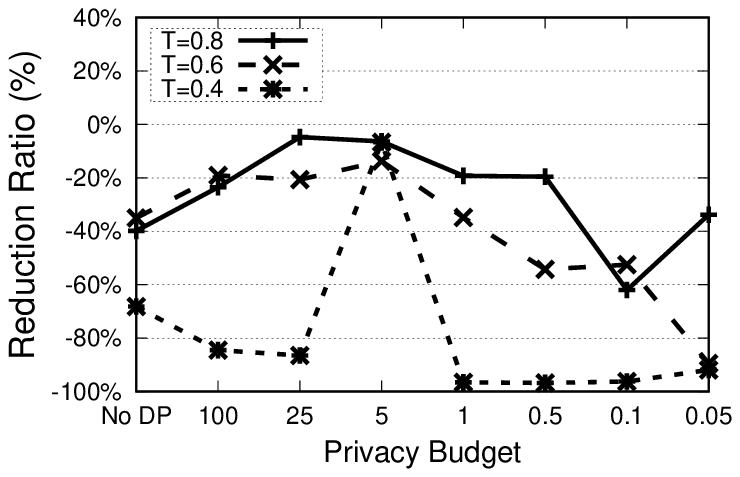}
          &
          \includegraphics[width=0.25\textwidth]{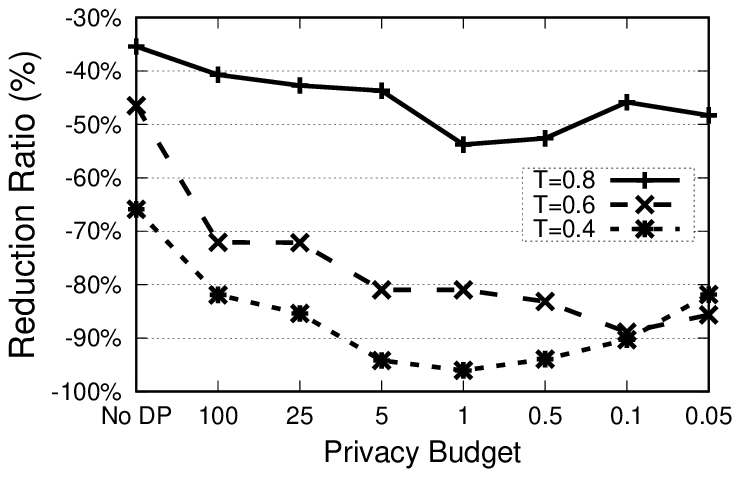}
           &
          \includegraphics[width=0.25\textwidth]{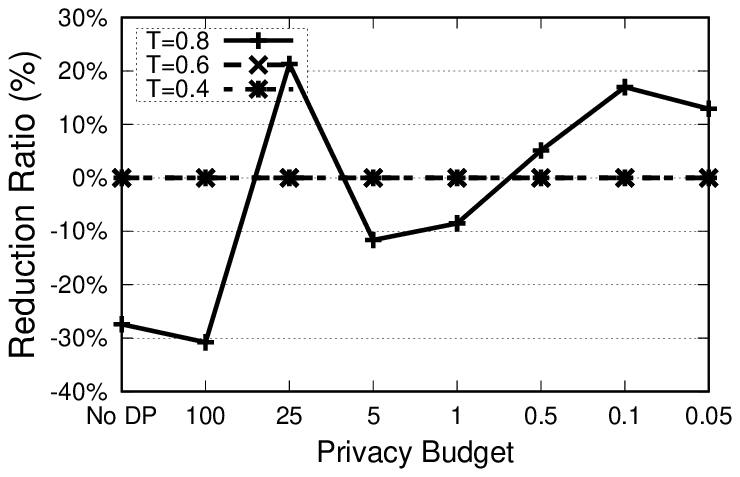}
        \\
        {\scriptsize (a) ADULT (Race)}
        &
        {\scriptsize (b) BROWARD (Race)}
         &
        {\scriptsize (c) COMPAS (Race)}
         &
        {\scriptsize (d) Hospital (Race)}
        \end{tabular}
        \vspace{-0.1in}
       \caption{\small \label{fig:real-reduc-gender} Effectiveness of \tool{} on VD reduction. Herein {\bf -x\%} indicates \tool{} reduces VD by {\bf x\%}.}
\end{figure*}

\noindent{\bf Challenges in practice.} Under practice settings, \tool{}, however, can encounter several challenges. We elaborate on these challenges and how we handle them in the following. 

\begin{itemize}[leftmargin=*]
\item As aforementioned, the distribution of feature combination is often sparse. Many feature combinations can only have 1 or 2 instances in a single group. For such feature combinations, \tool{} often gives small decimal values as the number of instances to delete. By intuition, we can simply round the decimal values up/down to integers. However, rounding these values up can result in data over-deletion while rounding them down will eliminate the reduction effects. To systematically address this problem, we propose \emph{feature aggregation}. Specifically, we run {\tt K-means} to classify the training samples into $K$ clusters and consider each cluster center as a new unique feature combination. When applying \tool{}, we replace each sample with its {\tt K-means} center. If data deletion is needed for a {\tt K-means} center, we randomly pick samples from the corresponding cluster. After the process of \tool{}, we then map the remaining samples back to their original values. The rationale behind our feature aggregation is that samples in the same cluster are in principle close and therefore, have similar probability of being attacked. To determine $K$ for feature aggregation, we pick the largest $K$ that ensures every group has at least a certain amount of samples in each cluster. 

\item In certain cases, \tool{} can require to delete a negative number of samples (\emph{i.e.}, $del(G_i, C_j) < 0$). This indicates the needs of duplicating samples. However, as is previously explained, sample duplication provides no real protection. To address this issue, we restrict the number of negative deletions. Specifically, we enumerate $T$ to find the values that preserves the utilities of the target model but requires less than an expected amount of negative deletions. With the above condition holds, we then simply ignore all the negative deletions requested by \tool{}.

\item In our explanation with the effects of \tool{} on VD, we assume the unsafe areas do not shift. This is, however, not the case in practice. As \tool{} alters the distribution of training samples, it will bring changes to the target model and hence, affect the unsafe areas. We argue that the shifting of unsafe areas should be insignificant because, as unveiled by the literature~\cite{long2018understanding}, MIA essentially attacks out-liers which usually span a specific ranges. Further, our evaluation in Section~\ref{subsec:eval} illustrates \tool{} works as exacted in practical settings where the unsafe areas may indeed shift.
\end{itemize}

To sum up, \tool{} follows Algorithm~\ref{alg:delete-reweigh} to reduce distribution difference among differen groups. 

\subsection{Evaluation of Unfairness Correction}
\label{subsec:eval}
To understand the utility of \tool{} for VD reduction, 
we perform a group of evaluation. In the following, we cover the details. 

\noindent{\bf Experiment setup.} In the evaluation, we first apply \tool{} to the training data 
and then follow Section~\ref{sc:dp-agaist-mia} for the reproduction of MIA and the use of 
DP against MIA. As introduced in Section~\ref{subsc:mia}, MIA trains label-class-specific 
attack models. Therefore, we group the training samples based on their labels and 
separately apply \tool{} to each of the sub-groups. Following Section~\ref{sc:vdmeasurement}, 
we consider gender and race as the protected groups (gender defines male and female; 
race defines black and non-black). Recall that \tool{} performs 
{\tt K-means} for feature aggregation. In this evaluation, we pick $K$ to ensure
each cluster has more than 10 samples. Further, we follow the approach described in 
Section~\ref{subsec:fairpick} to determine the threshold for VD reduction. 
 For the simplicity of presentation, we pick three thresholds that work for all 
the four data-sets (including {\tt 0.4}, {\tt 0.6}, and {\tt 0.8}). Finally, to 
reduce randomness, we repeat each test 25 times and report the average results. 

\noindent{\bf Impact on target model and MIA.} By intuition, \tool{} works by deleting 
training samples, which can hurt the target model and further affect 
the attack models. To verify this intuition, we compare the performance of the target model
and the attack models before and after we apply \tool{}. In Figure \ref{fig:target-model-acc-reweigh}, 
we present the comparison results. Note that to better visualize the comparison, 
we only show \tool{} with $T=0.8$. \tool{} with  $T=0.4$ and  $T=0.6$ has similar 
results. 

Overall, \tool{} can largely preserve the accuracy of the target model. In the case 
of Broward, the target model has nearly consistent accuracy after we apply \tool{}. 
On the data-sets of Adult and Hospital, \tool{} barely affects the target model's testing 
accuracy. It, however, slightly increases the training accuracy of the target model. 
This is most likely because, to reduce distribution difference, \tool{} deletes out-lier samples 
that only appear in certain groups and thus, makes the target model more centralized to better 
fit the training samples. In the case of Compas, \tool{} has no obvious impacts on the testing 
accuracy but increases the training accuracy by 15\% when we consider gender as the protected group. 
We believe this again is because \tool{} deletes irregularly-distributed samples such that the 
the target model becomes less scattered. As we can also observe in Figure 
\ref{fig:target-model-acc-reweigh}, \tool{} has similar effects on the target model 
with or without DP. This further proves that \tool{} is not impeding the target model \
and more importantly, \tool{} is compatible with using DP as a defense. 

With regards to the impacts on MIA, \tool{} shows similar patterns to its effects on the 
target model. This is understandable since MIA is largely determined by the target model. 
To be specific, on the Adult, Broward, and Hospital data-sets, MIA all has highly similar 
recall and precision with or without \tool{}. The only significant difference incurred 
by \tool{} is MIA's recall on the Compas data-set. When we consider gender as the protected 
attribute for Compas, \tool{} increases MIA's recall by around 10\%. This, as explained above, 
is because the data deletion by \tool{} makes the target model over-fit the training samples and 
therefore, leads MIA to have higher effectiveness. If we consider race as the protected 
attribute for Compas, MIA' recall is down-graded by around 10\%.
We believe this is likely due to randomness as the recall consistently increases when DP is augmented. 
To sum up, \tool{} only incurs slight (near zero in many cases) changes to the performance 
of MIA. This is critical for our study because this helps confirm that VD reduction (if observed)
is indeed attributable to \tool{} instead of variance in MIA's performance. 

\noindent{\bf Effectiveness of VD reduction.} In Figure~\ref{fig:real-reduc-gender}, 
we present the reduction of VD after we apply \tool{}. In general,
\tool{} demonstrates expected effects on VD reduction. Particularly, when the original VD 
is significant ($VD > 0.1$, Broward and Compas in Figure~\ref{fig:vul-disp}), 
we observe stable and effective reduction. With the two data-sets, the average reduction 
rate consistently exceeds 30\% (under different $T$). Also observable in Figure~\ref{fig:real-reduc-gender}
is that the reduction by \tool{} is independent with DP. This, again, proves that 
\tool{} is compatible with DP and can complement DP to provide fair protection. 

From Figure~\ref{fig:real-reduc-gender}, we can also see a pattern 
(despite somewhat unstable) that the reduction of VD increases while we decrease $T$
(namely increase the expected reduction rate). Such patterns are especially significant in 
the case of Compas (with race), Adult (with gender), Broward (with race), 
and Hospital (with gender). This shows that our set-up of $T$ works as expected, 
providing user the flexibility of choosing fairness level on demands. 

We also note cases where \tool{} does not reduce VD but instead increases its value. 
We reason each of the cases as follows. In the case of Adult (with gender as the protected
attribute) and Hospital (with race as the protected attributed), the original 
VD is very insignificant. A slight change or randomness will lead to 
major VD changes. As such, the effectiveness of \tool{} can be masked and 
not observable. In the case of Broward with gender as the protected attribute, 
\tool{} increases VD when $T=0.4$. In this case, among all the deletion requests by
\tool{}, over half of them are negative deletions (namely duplication). Recall that \tool{}
ignores such requests. Therefore, it has no reduction effects. On the contrary, its 
half-done reduction accidentally increases the variances in data distribution. 

To sum up, our evaluation shows \tool{} can provide reliable and effective VD reduction 
under practical settings. 

\vspace{-0.1in}
\section{Related Work}
\label{sc:related}

In this section, we review the related work on MIA attacks and defenses, DP and its evaluation, and fairness in ML. 

\noindent{\bf Membership inference attacks and defenses} MIA was initially proposed by Shokri \etal~\cite{shokri2017membership}. Under a black-box setting where the target model is unknown, MIA predicts whether a given record was used in training the target model. Afterwards, several follow-up works provide more detailed study. \cite{truex2019demystifying} characterized the attack vulnerability with respect to types of learning models, data distribution, and transferablity. 
\cite{salem2018ml} proposed new membership inference attacks by relaxing the assumptions
of the original attack in \cite{shokri2017membership} in both model types and data. 
\cite{long2018understanding} generalized MIA by identifying vulnerable records and indirect inference. 
Recently, \cite{nasr2019comprehensive} proposed new MIA attack against white-box ML models. 
Also MIA attack has been used to attack Federated learning \cite{nasr2019comprehensive}, collaborative learning \cite{melis2019exploiting}, generative adversarial networks (GANs) \cite{hayes2019logan}, and adversarially robust deep learning models \cite{song2019membership}.

Several defense mechanisms have been designed to defend against MIA, including dropout and model stacking~\cite{salem2018ml}, adversarial regularization \cite{nasr2018machine}, $L_2$-regularizer \cite{shokri2017membership}, and adversarial examples~\cite{jia2019memguard}. 

None of these techniques can provide a provable privacy guarantee against MIA. 
In this paper, we mainly consider differential privacy \cite{dwork2006calibrating} as the defense mechanism against MIA, as it can provide a theoretical privacy guarantee.

\noindent{\bf Differential privacy and Machine Learning}
Differential privacy (DP) ~\cite{dwork2006calibrating} has become the de facto standard in measuring the disclosure of privacy pertaining to individuals. To accommodate different types of machine learning models, there have developed various mechanisms to enforce DP~\cite{dwork2008differential,mcsherry2007mechanism,warner1965randomized}.
The mechanisms can be categorized into two types: (1) {\em pre-processing}: the training data is perturbed by generating synthetic data with differential privacy \cite{cormode2011differentially,chen2015differentially}; and (2) {\em in-processing}: add noise to the objective function  of the learning model (e.g., \cite{iyengar2019towards,chaudhuri2011differentially,zhang2012functional,jagannathan2009practical}) or 
or to the gradient in each iteration of gradient descent or stochastic gradient descent that is used
to minimize the objective function \cite{song2013stochastic,bassily2014private}. We refer the authors to some good surveys \cite{dwork2010differential,ji2014differential,jayaraman2019evaluating} for more readings. Going beyond classic machine learning models, Shokri \etal extend DP to deep neural networks~\cite{shokri2015privacy}. Following that, Abadi \etal propose an alternative approach that performs randomized perturbation during the stochastic gradient descent process \cite{abadi2016beckman}.
Unlike these works that focus on exploring DP mechanisms on machine learning models and strive for the balance between utility and privacy guarantees,  our research considers the fairness issues behind DP and endeavors to providing mitigation.

\noindent{\bf Algorithmic fairness in ML}  Several competing notions of fairness have been recently proposed in the machine learning literature.  
The definition of fairness can be categorized into three types \cite{quadrianto2017recycling}: 
i) it is not based on protected attributes such as gender or race ({\em fair treatment}), ii) it does not disproportionately benefit or hurt individuals ({\em fair impact}), and iii) given the target outcomes, it enforces equal discrepancies between decisions and target outcomes across groups of individuals based on their protected characteristic ({\em fair supervised performance}).

An example of fair treatment is fairness through unawareness \cite{grgic2016case} that ignores the protected attributes. 

Examples of fair impact constraints include 80\% rule \cite{feldman2015certifying} and demographic parity \cite{calders2009building,kamishima2012fairness}. Examples of fair supervised performance constraints include equal opportunity and equal odds \cite{hardt2016equality} and de-correlation \cite{zafar2017fairness}. Most of these definitions focus on fairness of groups (i.e., individuals who share the same value on the protected attributes). Individual fairness \cite{dwork2012fairness,vapnik2009new,kusner2017counterfactual} is defined  as a non-preferential treatment towards an individual. Counterfactual fairness \cite{kusner2017counterfactual,garg2019counterfactual} evaluates fairness in terms of causal inference and counterfactural examples. 

Techniques to design bias mitigation algorithms  typically identify a fairness notion of interest first, and modify a particular point of ML pipeline to satisfy it. Methodologically, they fall broadly into three categories: (1) {\em pre-processing}: the bias in the training data is mitigated \cite{calders2009building,kamiran2009classifying,feldman2015certifying}; (2) {\em in-processing}: the machine learning model is modified by adding fairness as additional constraint \cite{calders2010three,zafar2017fairness,goh2016satisfying}; and (3) {\em post-processing}:  the results of a previously trained classifier are modified to
achieve the desired results on different groups \cite{hardt2016equality}.

\vspace{-0.1in}
\section{Conclusion and Future Work}
\label{sc:conclusion}

Fairness and privacy are two equally important issues for machine learning. Most of the recent studies have investigated these two issues separately. No attention has been paid to fair privacy, i.e., privacy models and their enforcement mechanisms should not disproportionately benefit or hurt individuals. We are the first to examine fair privacy in the context of membership inference attack and differential privacy as the defense mechanism. We provide  extensive empirical evidence that vulnerability disparity against MIA exists, without and with DP applied. We performed detailed analysis to identify the source of such vulnerability disparity. Based on our findings, we designed a new mitigation method named \tool\ that adjusts the distribution of the training data. Our results show that \tool\ can effectively reduce VD for both without and with DP deployment. 

\noindent{\bf Future Work.} There are quite a few interesting research directions to explore. First, we will investigate if vulnerability disparity exists for different target models with DP (e.g., differentially private DNN \cite{abadi2016deep}) and different defense mechanisms (e.g., \cite{salem2018ml,nasr2018machine,shokri2017membership}) against MIA. Second, we will consider  different fairness metrics to evaluate vulnerability disparity. In this paper, we mainly focus on  vulnerability disparity of different groups. Recently, {\em individual fairness} has attracted much attention in the machine learning community. Briefly speaking, individual fairness requires that similar objects should receive similar treatment. Consider DP defends against MIA.  If a record $R_1$ is exposed by MIA but its similar records are not, apparently it does not ensure the individual fairness. 
We will examine if such individual vulnerability disparity against MIA exists before and after DP. 
As pointed out by \cite{long2018understanding}, an outlier is more likely to be a vulnerable target record. 
This can be the starting point of the analysis of individual vulnerability disparity.

\bibliographystyle{ACM-Reference-Format}
\bibliography{main}
\end{document}